%% file: main.tex
\documentclass{article}

\usepackage[english]{babel}

\usepackage[letterpaper,top=2cm,bottom=2cm,left=3cm,right=3cm,marginparwidth=1.75cm]{geometry}

\usepackage{amsmath}
\usepackage{graphicx}
\usepackage{authblk}
\usepackage{multirow}

\usepackage[colorlinks=true, allcolors=blue]{hyperref}

\title{Hybrid pattern recognition for charged particle tracking: \Hough transform and convolutional neural efficiency networks}

\author[1]{Carlo Varni}
\author[1]{Krzysztof Cie\'sla}
\author[2]{Marcin Wolter}
\author[1]{Radoslaw Karol Mie\'nkowski}
\author[3]{Noemi Calace}
\author[1]{Tomasz Bo\l d}

\affil[1]{AGH University of Krakow, Poland}
\affil[2]{Institute of Nuclear Physics Polish Academy of Sciences, Krakow, Poland}
\affil[3]{CERN, Geneva, Switzerland}

\date{}

\usepackage{graphicx}
\usepackage{subcaption}

\input{commands}

\usepackage{lineno}

\begin{document}
\maketitle

\begin{abstract}

\input{sections/abstract}
\end{abstract}

\section{Introduction}
\label{sec:introduction}

\input{sections/introduction}

\section{Simulated data}
\label{sec:simulated_data}

\input{sections/simulated_data}

\section{Algorithm description}
\label{sec:algorithm_description}

\input{sections/algorithm_description}

\section{Performance}
\label{sec:performance}

\input{sections/performance}

\section{Parallelization capabilities}
\label{sec:parallelization_capabilities}
\input{sections/parallelization_capabilities}

\section{Summary and outlook}
\label{sec:summary}

\input{sections/summary}

\section*{Declaration of generative AI and AI-assisted technologies in the manuscript preparation process}
\label{sec:ai}
\input{sections/ai}

\section*{Acknowledgements}
\label{sec:acknowledgements}
\input{sections/acknowledgements}

\bibliographystyle{unsrt}
\bibliography{cas-refs}

\end{document}

%% file: commands.tex
\usepackage{xspace}

\newcommand{\ACTS}{ACTS\xspace}
\newcommand{\Hough}{Hough\xspace}
\newcommand{\pileup}{pile-up\xspace}

%% file: sections/abstract.tex
Reconstructing charged-particle tracks in silicon detectors is a central task in high-energy physics experiments and a key component of both offline reconstruction and online event selection.
Within the reconstruction chain, the efficient and high-purity formation of track candidates plays a critical role in the overall performance.
Among the many approaches developed over the years, the \Hough transform (HT) has been widely studied as a fast geometry-driven method for track finding. However, in high-occupancy environments such as those expected at the High-Luminosity LHC (HL-LHC), the HT tends to produce a large number of spurious candidates, leading to increased computational overhead in subsequent reconstruction stages.
In this work, we present a hybrid approach in which the HT serves as a first-stage data preparation step, providing its parameters space image as an input to a neural network trained to suppress false track candidates. The method combines the speed of the HT with the discriminative power of machine learning to achieve both efficiency and purity. In addition no data transformations are involved when combining these steps resulting in a simpler and more performant algorithm. Performance studies using the Open Data Detector simulated in the \ACTS framework under realistic HL-LHC \pileup conditions will be presented.

%% file: sections/introduction.tex
The reconstruction of charged-particle trajectories is a central component of event reconstruction in high-energy physics experiments. However, the increasingly complex detector geometries and high-occupancy collision environments anticipated at future facilities pose significant challenges to track reconstruction. These challenges arise from the inherently combinatorial nature of pattern recognition, as the number of possible hit combinations grows rapidly with detector occupancy.
This issue is particularly pronounced in track seeding, which identifies groups of hits compatible with an helix trajectory as starting points for the subsequent iterative pattern-recognition procedure. Traditional seeding approaches typically exhibit quadratic or even cubic scaling with the number of detector hits and layer combinations considered. Furthermore, they often generate a large number of spurious candidates. Although incorporating additional detector information can help suppress fake seeds, it also increases the combinatorial complexity by introducing a larger number of possible hit combinations.

High-energy physics experiments~\cite{ATLAS_Collaboration_2008, CMS_Collaboration_2008} -- such as those at the Large Hadron Collider (LHC)~\cite{lhc} -- already operate at very high instantaneous luminosities, but the upcoming High-Luminosity LHC (HL-LHC)~\cite{hl-lhc} era is expected to further push the event complexity, with an expected number of simultaneous events (\pileup, $\mu$) up to 200.  In this environment, track reconstruction will represent the dominant computational challenge in data processing~\cite{ATLAS-TDR-PHASE2, CMS-TDR-PHASE2}. 
The development of novel approaches is pivotal: algorithms whose complexity scales linearly -- or near-linearly -- with the number of measurements are therefore attractive due to their favourable scaling properties.
Moreover, while central processing units (CPUs) are the current technology of choice for data processing, heterogeneous computing architectures are becoming increasingly relevant. As the adoption of computational accelerators such as graphics processing units (GPUs) and field-programmable gate arrays (FPGAs) is being considered -- particularly for real-time tracking applications -- there is growing demand for algorithms that are naturally suited for such architectures in terms of parallelization capabilities.

An effective seeding algorithm must balance high efficiency with a low rate of fake combinations. 
A controlled level of redundancy — i.e. multiple seeds corresponding to the same charged particle — is desirable to ensure robustness against local inefficiencies and detector effects.
Track seeding is typically performed by combining detector measurements under a set of physical and geometrical constraints: compatibility with a known interaction point; curvature induced by the magnetic field; and expected multiple scattering from material interactions. The method of choice for the subsequent track finding in most experiments remains the Kalman filter~\cite{Kalman1960, Fruhwirth1987}. However, novel approaches based on graph-based machine learning techniques have been investigated~\cite{ju2020graph}.

Another prominent example is the \Hough transform (HT)~\cite{duda1972use}, well known from computer vision applications. The HT has been successfully applied to charged particle tracking in several contexts~\cite{Alfonsi_2024}. However, when used for track reconstruction in central detectors operating at high luminosities, the method tends to produce a significant number of false positives.

In this study, we pursue the idea of post-processing the output of the \Hough transform with a compact machine learning model trained to retain valid track seeds with high efficiency while rejecting fake candidates. The key novelty of this approach is that, once the \Hough accumulator is constructed from detector measurements, the subsequent machine learning inference is performed directly in image space, avoiding costly data transformations. Since the algorithm operates on image-like representations, efficient implementations on heterogeneous hardware architectures -- particularly GPUs -- are naturally supported~\cite{chen2011accelerating}.

This document is structured in the following fashion: a first description of the detector and the infrastructure for producing the data used in this study is presented in section~\ref{sec:simulated_data}. A description of the algorithm and workflow is described in section~\ref{sec:algorithm_description}, followed by a showcase of the performance on section~\ref{sec:performance}. An overview of the paralellization possibilities of the proposed algorithm will be discussed in section~\ref{sec:parallelization_capabilities}.

%% file: sections/simulated_data.tex
The study is based on simulated data produced using the \ACTS (A Common Tracking Software) software~\cite{acts-2022}.
\ACTS is an experiment- and framework-independent toolkit for charged-particle track reconstruction that provides a comprehensive set of high-level tracking algorithms and utilities through its core library. This library contains the main building blocks for detector geometry description, track finding and fitting, and event reconstruction.
In addition, \ACTS is distributed with an examples framework: a stand-alone environment designed for algorithm R\&D, validation, and performance studies. The examples framework provides complete simulation and reconstruction workflows, enabling users to generate events, simulate detector responses, reconstruct tracks, and evaluate new algorithms in a realistic setting. Event generation can be performed either with a simple \ACTS particle gun or with Pythia8~\cite{SJOSTRAND2008852, pythia}, while detector simulation is based on Geant4~\cite{AGOSTINELLI2003250}.

\begin{figure*}
    \centering
    \includegraphics[width=0.95\textwidth]{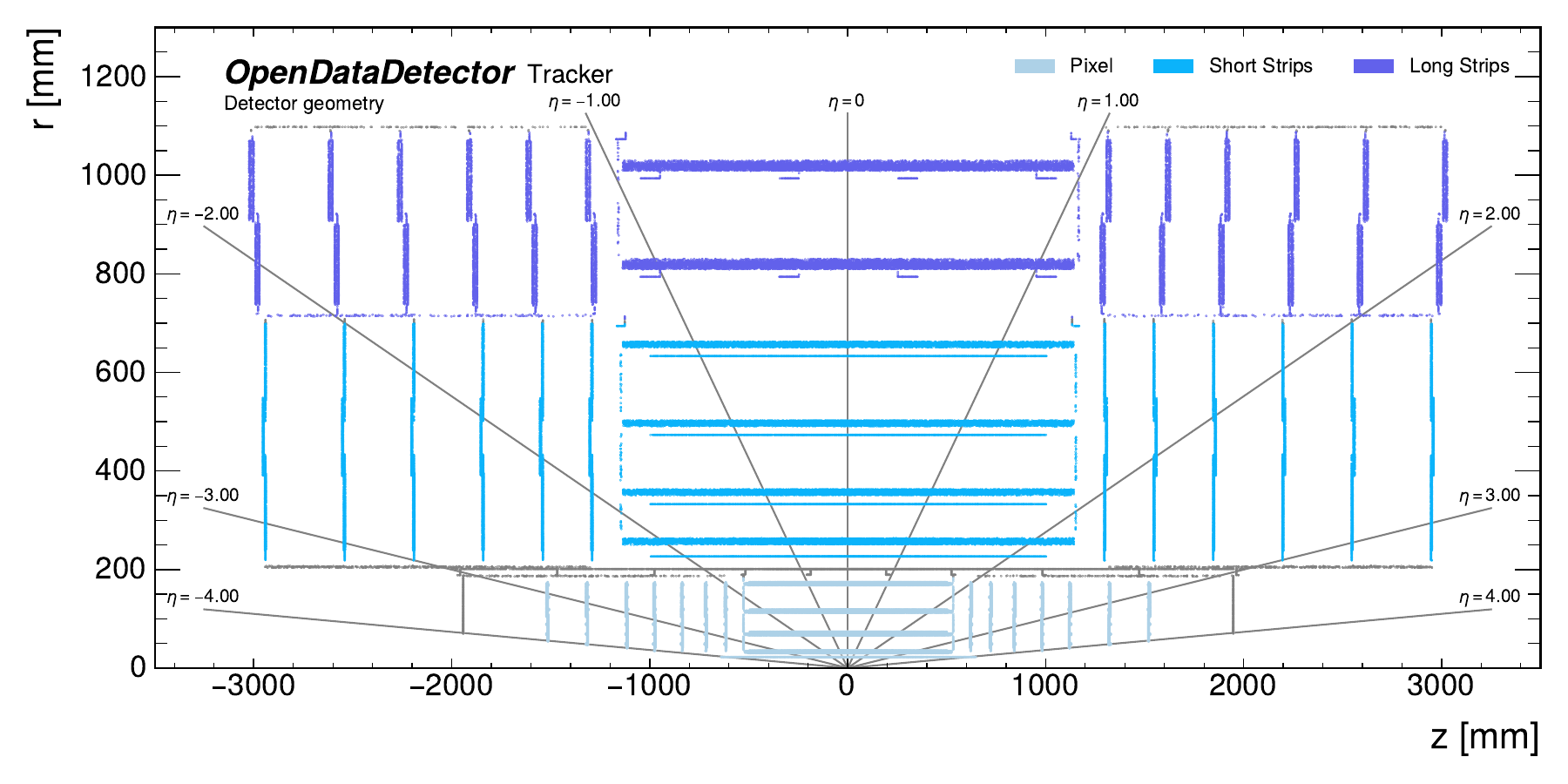}
    \caption{Layout of the Open Data Detector's tracker, segmented into its sub-components: Pixels; Short Strip; and Long Strips. Layers and disks are represented by solid lines. Pseudorapidity values are shown.}
    \label{fig:ODDdetectorLayout}
\end{figure*}

\ACTS also comes with an open-to-the-public and free-of-license detector geometry, which is used for developing and validating algorithms: the Open Data Detector (ODD)~\cite{gessinger2023open}. It contains a tracker detector as well as calorimeters and muon chambers.
The ODD is an evolution of the detector geometry used for the Tracking Machine Learning (TrackML) challenge~\cite{trackML1, trackML2, trackML3} and its tracker is an LHC-like full silicon prototype inner detector based on DD4hep~\cite{dd4hep}. The layout of the ODD tracker is shown in figure~\ref{fig:ODDdetectorLayout}, taken from reference~\cite{elitez2025collidermlreleaseopendatadetectorhighluminosity}. It consists of 3 sub-detectors: Pixel, Short Strips, and Long Strips. The ODD tracker is designed to provide at least 12 layers of sensitive material in the range $|\eta| \leq 3.0$, and 8 layers for $|\eta| \leq 3.5$. Details of these sub-detectors are shown in table~\ref{tab:ODDdetectorInfo}.
The ODD detector has been used to produce in the recently-released ColliderML dataset~\cite{elitez2025collidermlreleaseopendatadetectorhighluminosity}.

\begin{table*}[htbp]
\centering
\begin{tabular}{|l|c|c|c|c|}
\hline
\textbf{Sub-detector} & \textbf{Measurement} & 
\begin{tabular}[c]{@{}c@{}}\textbf{Barrel}\\\textbf{pitch}\end{tabular} &
\begin{tabular}[c]{@{}c@{}}\textbf{End-cap}\\\textbf{pitch}\end{tabular} &
\begin{tabular}[c]{@{}c@{}}\textbf{\# Barrel Layers}\\\textbf{/ \# Endcap Disks}\end{tabular} \\
\hline
Pixel & 2D (+time) & $50\times50$~$\mu m^{2}$ & $50\times50$~$\mu m^{2}$ & 4 / 7 \\
\hline
Short Strip & 2D & $80\times500$~$\mu m^{2}$ & $80\times500$~$\mu m^{2}$ & 4 / 6 \\
\hline
Long Strip & 1D & 100~$\mu m$ & 125~$\mu m$ & 2 / 6 \\
\hline
\end{tabular}
\caption{Overview of the Open Data Detector tracking sub-detectors. The table summarizes the measurement dimensionality, sensor pitch in the barrel and end-cap regions, and the number of barrel layers and end-cap disks for the pixel, short strip, and long strip detectors.}
\label{tab:ODDdetectorInfo}
\end{table*}

The samples used for this study comprise:
\begin{itemize}
    \item Single-muon events generated with the \ACTS particle gun with $\langle\mu\rangle = 0$. These are generated in the $[1, 100]$~GeV $p_{T}$ range and within the pseudorapidity range of $|\eta| \leq 3$.
    \item Single-muon events generated with the \ACTS particle gun with $\langle\mu\rangle = 200$. These are generated in the $[1, 100]$~GeV $p_{T}$ range and within the pseudorapidity range of $|\eta| \leq 3$.
    \item $t\bar{t}$ events generated with Pythia8 with $\langle\mu\rangle = 200$ at a center of mass energy of $\sqrt{s} = 14$~TeV.
\end{itemize}

Detector simulation is performed with Geant4. A smeared digitization approach is used to simulate the detector response. Simulated hit positions are smeared using Gaussian noise corresponding to the detector resolution of the sub-detectors.

Single-muon events without \pileup are used as a baseline sample for algorithm validation and performance characterization. These events provide a clean and controlled environment in which intrinsic algorithmic properties, such as seeding efficiency and fake rate, can be studied without the confounding effects of high detector occupancy. 
Single-muon events with \pileup are used to provide the neural network with a controlled source of high-$p_T$ tracks embedded in high-occupancy environments. This complements the predominantly low- to moderate-$p_T$ track spectrum of the $t\bar{t}$ sample, which will also be used in the neural network training procedure.
A second, independent $t\bar{t}$ sample will be used for performance evaluation. 
This dataset reflects the challenging conditions expected at high-luminosity hadron colliders -- characterized by a large number of simultaneous interactions and dense hit environments --, and will enable a realistic assessment of the robustness, scalability, and suitability for HL-LHC conditions of the algorithm.

%% file: sections/algorithm_description.tex
The track reconstruction strategy presented in this work is based on an extended formulation of HT for the seed finding process, specifically designed to operate robustly in high-occupancy environments and under conditions of elevated track density, as expected in future high-luminosity scenarios.
The input to the HT-based seeding algorithm consists of space points, i.e. three-dimensional points corresponding to particle signals in the detector. 
Space points are constructed from hits in the pixel and short strip detectors by applying a local-to-global coordinate transformation to the measured clusters. For the long strip detector, space points are formed by combining clusters from opposite sides of a module under the assumption that the particle trajectory originates from the interaction region.

\begin{figure*}
    \centering
    \includegraphics[width=0.95\textwidth]{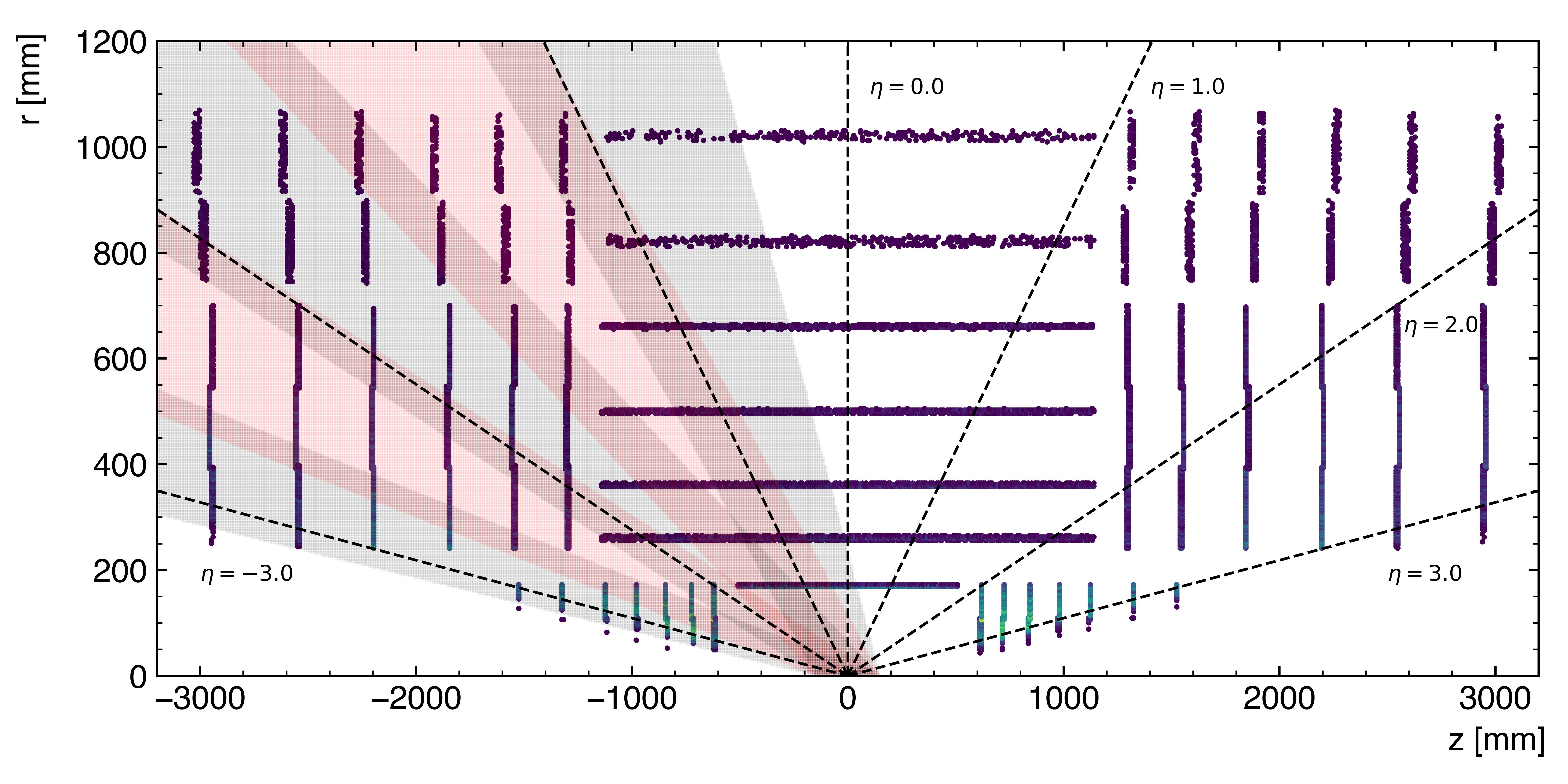}
    \caption{Visualization of the sub-division of the detector pseudorapidity coverage into $\eta$ slices. Only few slices are shown. The hits selected by all slices, and used in this study, are also shown.
    }
    \label{fig:event_display}
\end{figure*}

The algorithm exploits a subdivision of the kinematic parameter space by dividing the detector pseudorapidity acceptance into a set of slices, which define Regions of Interest (RoI). The detector acceptance is partitioned into $13~\eta$ intervals that fully cover the pseudorapidity range. Studies showed that this number of slices is enough to provide a good physics performance and robustness. The definitions of these slices and the corresponding $\eta$ coverage are shown in table~\ref{tab:slice_definition}.
In order to suppress edge effects -- in particular for trajectories intersecting the boundary between adjacent slices -- the RoIs are constructed with a controlled overlap. This overlapping configuration ensures continuity of reconstruction performance across slice boundaries and prevents efficiency losses induced by the subdivision of the parameter space. The RoI definition further assumes that charged particles originate within a longitudinal interval $\Delta z = \pm 150$~mm around the nominal interaction point. This constraint reflects the expected spread of the primary vertices along the beam axis and is incorporated into the geometrical model underlying the slice construction.
The partitioning is shown in figure~\ref{fig:event_display}, with a few of the slices overlaid on top of the space points in the detector.

\begin{table*}[htbp]
\centering
\begin{tabular}{|c|c|c|c|c|c|}
\hline
\textbf{\# slice} & 
\begin{tabular}[c]{@{}c@{}}\textbf{$\eta$ coverage}\\\textbf{(center $\pm$ width)}\end{tabular} & 
\textbf{\# slice} & 
\begin{tabular}[c]{@{}c@{}}\textbf{$\eta$ coverage}\\\textbf{(center $\pm$ width)}\end{tabular} &
\textbf{\# slice} & 
\begin{tabular}[c]{@{}c@{}}\textbf{$\eta$ coverage}\\\textbf{(center $\pm$ width)}\end{tabular} \\
\hline

\textbf{0} & $-2.77 \pm 0.23$ & \textbf{5} & $-0.46 \pm 0.23$ & \textbf{10} & $+1.85 \pm 0.23$ \\
\hline
\textbf{1} & $-2.31 \pm 0.23$ & \textbf{6} & $+0.00 \pm 0.23$ & \textbf{11} & $+2.31 \pm 0.23$ \\
\hline
\textbf{2} & $-1.85 \pm 0.23$ & \textbf{7} & $+0.46 \pm 0.23$ & \textbf{12} & $+2.77 \pm 0.23$ \\
\hline
\textbf{3} & $-1.38 \pm 0.23$ & \textbf{8} & $+0.92 \pm 0.23$ & & \\
\hline
\textbf{4} & $-0.92 \pm 0.23$ & \textbf{9} & $+1.38 \pm 0.23$ &  & \\
\hline
\end{tabular}
\caption{Central values and coverage of the $\eta$ slices used in the study. The full acceptance is partitioned into 13 overlapping regions, each spanning approximately $\pm 0.23$ in $\eta$ around the central value listed in the table. The spread of the slices along the interaction region causes the overlap.}
\label{tab:slice_definition}
\end{table*}

Not all space points are used for the seed-finding process. For each $\eta$ slice, only the space points that satisfy the selection criteria of the slice are used. The space points from all the layers in the strip sub-detectors are used, as well as the space points from all the end-caps and from the last layer of the barrel of the pixel sub-detector.
Space points from the three innermost layers of the pixel barrel detector are not considered due to their high occupancy, which originates from the large hit density close to the interaction point.
A \Hough transformation is then applied to this subset of space points, projecting each point defined by radial coordinates $r_s, \varphi_s$ into a linear function in parameter space defined by the charge-signed inverse transverse momentum, $q/p_{T}$, and the azimuthal emission angle, $\phi$.
Under the assumption of helical motion in a uniform solenoidal magnetic field and small trajectory curvature, the transformation is defined by: 
\begin{equation}
   r_s, \varphi_s \rightarrow 
   \frac{q}{p_T} = \frac{\phi-\varphi_s}{r_s A}
\end{equation}
where $A$ is a constant that depends on magnetic field induction. 
This procedure produces two-dimensional histograms, referred to as the \Hough planes, in which each bin accumulates contributions compatible with a given helical trajectory in the detector. 
The \Hough plane is filled by counting the number of distinct detector layers contributing to each bin, rather than the total number of space points. This enhances the discrimination power between genuine track candidates -- which typically traverse multiple detector layers -- and random or noisy combinations originating from unrelated close by hits on the same layer.

\begin{figure*}
    \centering
    \begin{subfigure}{0.48\textwidth}
        \centering
        \includegraphics[width=\textwidth]{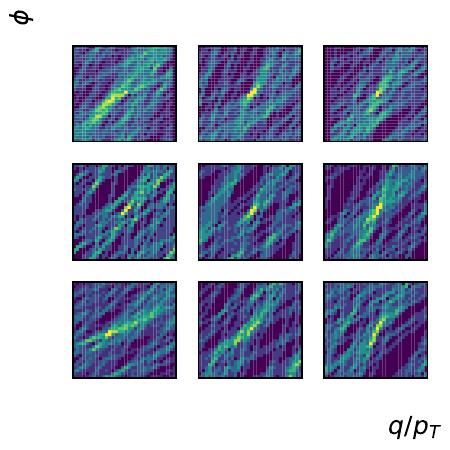}
    \end{subfigure}
    \hfill
    \begin{subfigure}{0.48\textwidth}
        \centering
        \includegraphics[width=\textwidth]{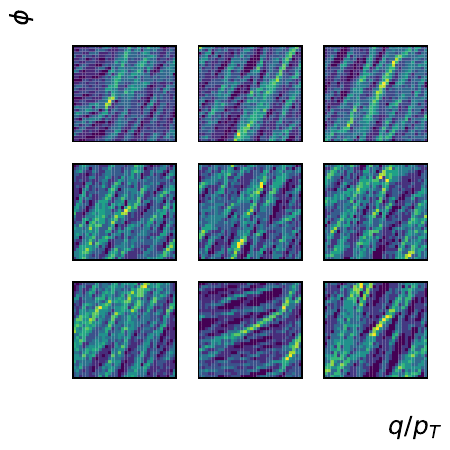}
    \end{subfigure}

    \caption{Train sample images from a mixture of $t\bar{t}$ and single muon samples with a $\langle \mu \rangle = 200$: (left) true peaks; (right) fake peaks.
    }
    \label{fig:ttBarWindows}
    
\end{figure*}

Once the \Hough plane has been filled, the histograms are analyzed to identify the peaks. The peak-finding procedure searches for local maxima in the $(q/p_{T}, \phi)$ planes. In cases where plateaux are observed -- i.e. regions of adjacent bins with comparable values that do not exhibit a single well-defined maximum -- a clustering procedure is applied to the peaks in order to group bins compatible with the same physical track and extract a representative candidate for further processing. For each peak, a fixed-size image of $32 \times 32$~bins around it is extracted from the \Hough plane. A bin size of $32 \times 32$ pixels was chosen as a compromise between computational efficiency and information retention: this resolution is large enough to preserve the main structural features and intensity patterns of the images while keeping the data dimensionality low, which reduces memory usage and accelerates processing and model training. 
Examples of such images are shown in figure~\ref{fig:ttBarWindows}.

The images provide a localized representation of the \Hough plane around each candidate and are used as input to a neural network classifier.
The neural network is trained to determine whether a given peak corresponds to a genuine truth particle or originates from combinatorial background or noise. 
The classification is based on the information contained in the $32 \times 32$ \Hough plane images, as well as additional information such as the $q/p_{T}$ position of the peak. 
Details of the neural network structure will be discussed in section~\ref{sec:neural_network_configuration}.
Based on the output of the neural network only the peaks identified as genuine track candidates are retained. 
The results from all processed $\eta$ slices are then collected and merged into a single set of validated \Hough peaks.
Each retained peak corresponds to a well-defined collection of space points contributing to the associated \Hough bin or clustered region. 
These space points are extracted and used to form track seeds. Thanks to the \Hough-based selection and the neural network filtering, the resulting seeds typically consist of more than three space points. 
These seeds are subsequently provided as input to the traditional Combinatorial Kalman Filter (CKF), which is not covered in this study. The CKF will use all the measurements and layer of the ODD tracker to reconstruct tracks.

An intrinsic characteristic of the algorithm is that duplicate candidates are produced: a single truth particle may give rise to multiple peaks in the same \Hough plane. The overlap between $\eta$ slices also contributes to the reconstruction of duplicate candidates across slices.
The resolution of such duplicates is delegated to a subsequent stage of the reconstruction workflow.

A key aspect of the implementation concerns the binning scheme adopted for the \Hough plane. 
The granularity of the \Hough plane plays a crucial role in the overall computational complexity of the algorithm, since the number of bins determines the size of the parameter-space histogram and therefore affects both memory requirements and processing time. While a reduction of the binning would be expected to improve computational performance, preliminary studies indicated that decreasing the number of bins in either dimension of the Hough plane results in a non-negligible deterioration of the reconstruction performance.
Since the scope of this work is limited to the evaluation of physics performance, no systematic optimization of CPU consumption was carried out. The nominal $256 \times 7000$ binning, for $q/p_T$ and $\phi$ respectively, was therefore adopted for all studies presented in this paper.
Moreover, the binning along the $q/p_{T}$ coordinate is non-uniform. A non-equidistant binning is implemented, featuring significantly finer granularity in the region corresponding to $p_{T} > 20$~GeV. 
The motivation for this choice is that high-$p_T$ tracks correspond to small curvature values and they occupy a compressed region of the $q/p_T$ dimension. Under uniform binning, this compression degrades peak separation and efficiency. 
This non-uniformity of the \Hough space binning introduces a discontinuity in bin width that can bias the peak-finding procedure toward larger bins. This effect manifests as an increased rate of duplicate candidates in the binning transition region.

\subsection{Neural network configuration}
\label{sec:neural_network_configuration}

\begin{figure*}
    \centering
    \includegraphics[width=0.95\textwidth]{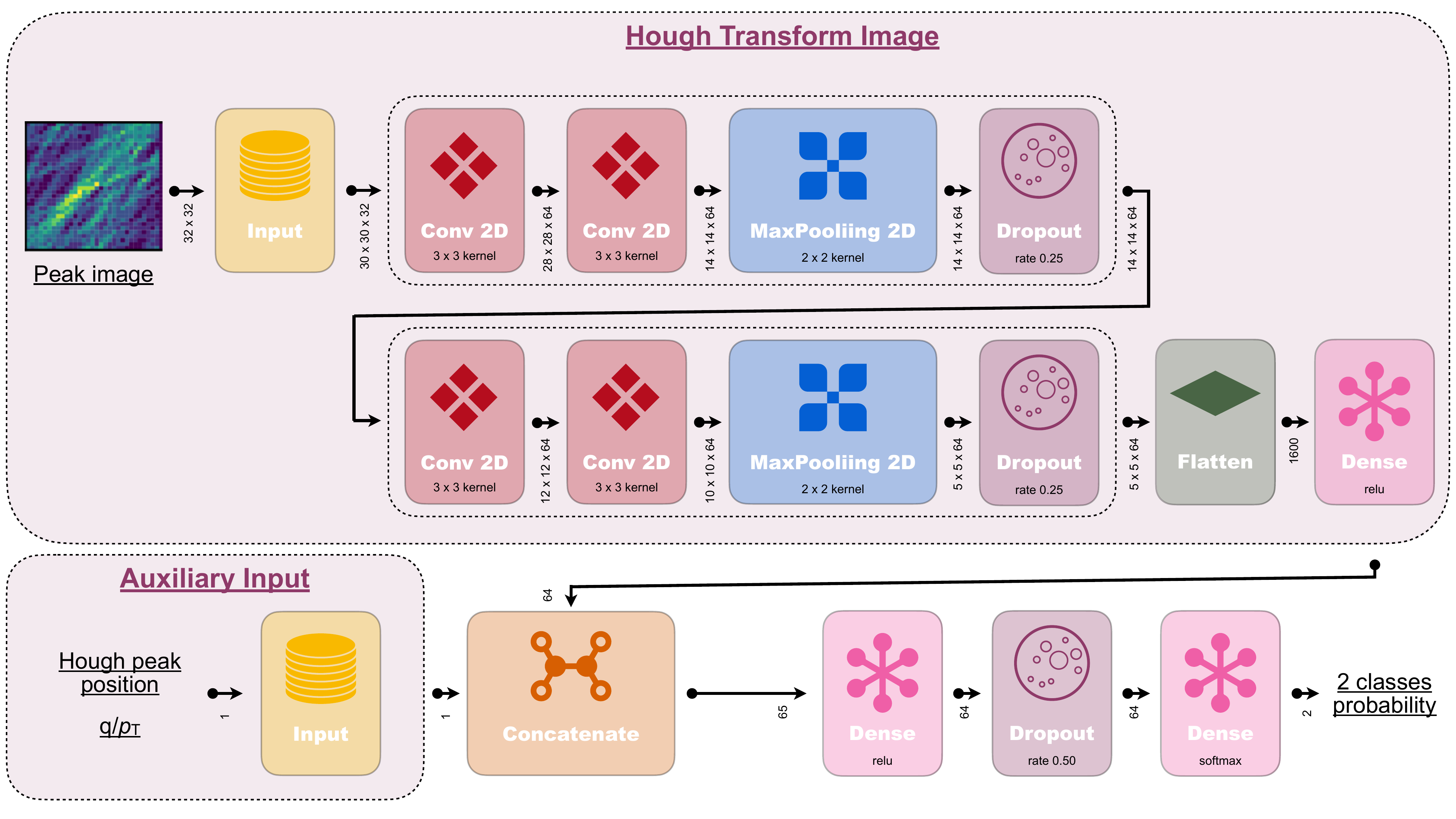}
    \caption{Schematic representation of the convolutional neural network internal architecture used in this analysis. Each layer is represented by a colored rectangle according to the type. Layer output tensors shapes and sizes are listed in between and characteristic layer parameters are mentioned within the boxes.}
    \label{fig:nnStructure} 
\end{figure*}

The baseline neural network used in this study is a compact convolutional neural network (CNN)~\cite{LeCun2015, NIPS2012_c399862d}, implemented with the TensorFlow~\cite{tensorflow2015-whitepaper} and Keras~\cite{chollet2015keras} frameworks. It is employed to discriminate genuine peaks, i.e. peaks associated with a truth particle, from fake peaks identified by the peak-finding algorithm. The structure of the baseline neural network is shown in figure~\ref{fig:nnStructure}.

The model has two inputs: a $32 \times 32$ image tensor; and an auxiliary metadata vector containing the $q/p_T$ position of the peak in the \Hough plane. The image branch consists of four convolutional layers with $3 \times 3$ kernels and Rectified Linear Unit (ReLU) activations, arranged in two Conv~$\rightarrow$~Conv~$\rightarrow$~MaxPool blocks. 
A dropout layer with rate 0.25 is applied after each pooling stage. 
The resulting feature maps are flattened and passed through a fully connected layer with 64 units to form a latent representation.
The auxiliary input is concatenated with the image-based representation. 
The combined feature vector is processed by a fully connected layer with 64 ReLU-activated units, followed by a dropout layer with rate 0.5 and a final softmax layer for classification.
To reduce potential biases arising from different population in the $q/p_T$ distributions of signal and background events, a re-weighting procedure is applied during training. 
The weights are computed separately for images in signal and background samples such that the $q/p_T$ distributions are flattened within each class. 
The resulting weights are normalized to unit mean and applied as per-sample weights in the loss function.
The network contains approximately $2\times10^{5}$ trainable parameters. 

\begin{figure*}
    \centering
    \begin{subfigure}{0.48\textwidth}
        \centering
        \includegraphics[width=\textwidth]{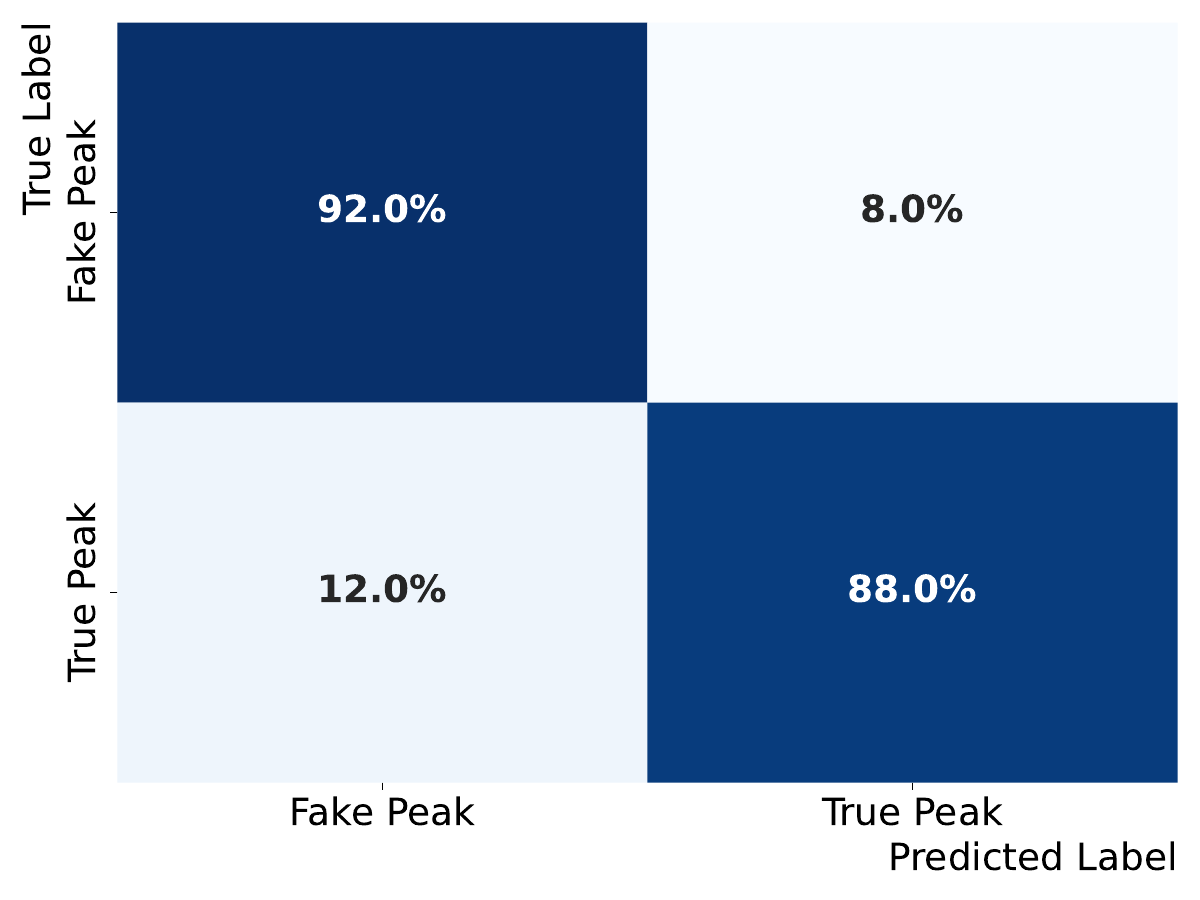}
    \end{subfigure}
    \hfill
    \begin{subfigure}{0.48\textwidth}
        \centering
        \includegraphics[width=\textwidth]{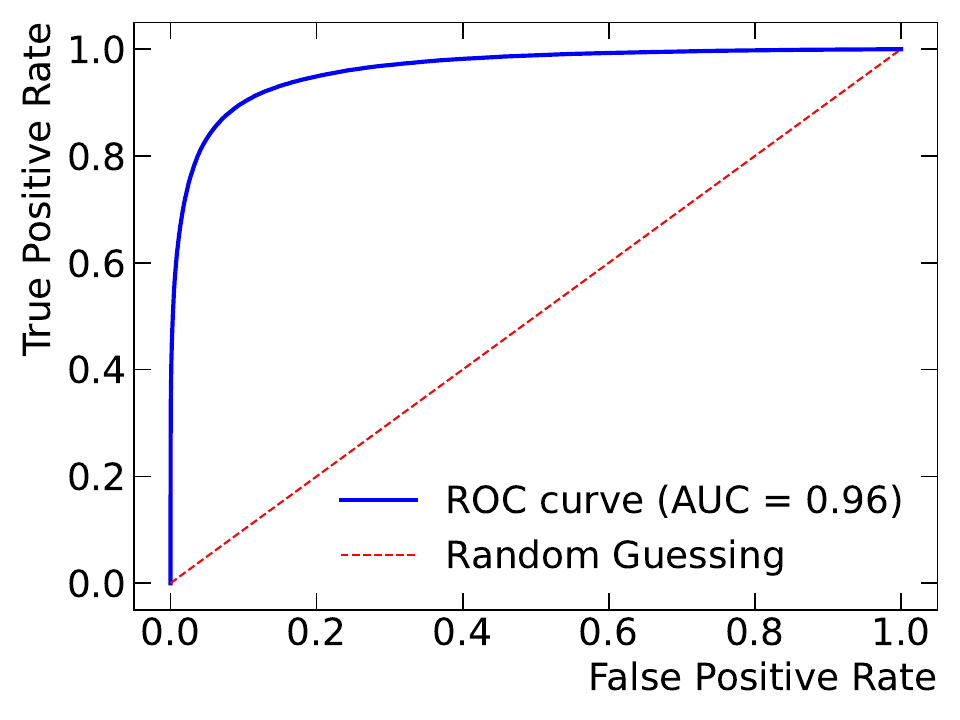}
    \end{subfigure}

    \caption{Performance of the neural network in selecting good track candidates: (left) the normalized confusion matrix; and (right) the ROC curve obtained from the neural network training.}
    \label{fig:nnPerformanceTraining}
    
\end{figure*}

Training is performed using a mixture of events containing single muons of  $p_T \in [1, 100]$~GeV and $t\bar{t}$ samples with $\langle \mu \rangle = 200$, providing coverage across the transverse-momentum spectrum. 
The former sample is used to assure population of statistics at the high values of $p_T$.
The performance of the network is evaluated using the confusion matrix and the Receiver Operating Characteristic (ROC) curve, as shown in figure~\ref{fig:nnPerformanceTraining}.

%% file: sections/performance.tex
The performance of the \Hough transform-based track finding algorithm is evaluated using both idealized and realistic simulation scenarios. 
The study begins with a simplified setup based on single-muon events without \pileup, which allows for the intrinsic performance of the algorithm to be assessed in the absence of additional detector activity. The analysis is subsequently exposed to a more challenging environment of $t\bar{t}$ events with $\langle\mu\rangle = 200$, representative of the conditions expected at the HL-LHC. This approach enables a systematic investigation of the \pileup dependence of the HT performance.

The system is designed to reconstruct track seeds for truth particles with transverse momentum $p_T \geq 1$ GeV.
The performance is evaluated for truth particles that satisfy the following selection criteria, to retain only those that have left a sufficient number of hits in the detector to be reconstructible:
\begin{itemize}
    \item Originates from any primary vertex within the interaction region $|vertex_z| \leq 150$~mm
    \item $p_{T} \geq 1$~GeV
    \item $|\eta| \leq 3$
    \item Number of hits left in the detector $\geq 7$
    \item Number of pixel hits $\geq 3$
    \item First pixel layer hit $\geq 1$
\end{itemize}

The performance of the HT is quantified using three complementary metrics: technical efficiency, duplicate rate, and fake rate. The technical efficiency is defined as the fraction of truth particles that are associated with at least one reconstructed peak. The duplicate rate is defined as the fraction of truth  particles that are associated with more than one reconstructed peak. The fake rate is defined as the fraction of reconstructed peaks that cannot be associated with any truth particle.
The association between reconstructed peaks and truth particles is established using information recorded during the filling of the \Hough planes: for each bin in the \Hough plane, the contributing truth particles are stored. 
A truth particle is associated with a bin if its hit contribution exceeds 50\% of the total bin content: 
\begin{equation}
    N^{pixel}_{hit} + N^{long\ strip}_{hit} + N^{short\ strip}_{hit}
\end{equation}

By construction, at most one truth particle can satisfy the 50\% threshold. In cases where multiple truth particles contribute to the same bin, only the particle satisfying these criteria is considered associated. If no particle satisfies this criteria, the bin is considered unassigned.
For each reconstructed peak, the truth association is evaluated using the \Hough plane bin corresponding to the peak position in the $(q/p_{T}, \phi)$ parameter space.

The performance of peak finding and the following CNN-based peak filtering stage is presented.
Also, to investigate the impact of the neural network architecture on the seeding performance, several increasingly compact CNN models are evaluated using the $t\bar{t}$ sample with $\langle\mu\rangle = 200$. 

Finally, peaks identified by the neural network as true peaks are used to generate seeds. These seeds are then processed by the CKF, which reconstructs the tracks. The track-finding performance is evaluated using a sample of $t\bar{t}$ events with a \pileup of 200 to assess the downstream impact of the HT procedure.
To quantify the reconstruction performance, the track reconstruction technical efficiency is computed and defined as the fraction of reconstructable truth particles that satisfy the above selection criteria and are successfully associated with a reconstructed track. A truth particle is associated with a track if its hit contribution exceeds 50\% of the total track content. 

\subsection{Seeding performance with single muons}
\label{sec:performance_with_single_muons}

\begin{figure*}
    \centering

    \begin{subfigure}{0.48\textwidth}
        \centering
        \includegraphics[width=\textwidth]{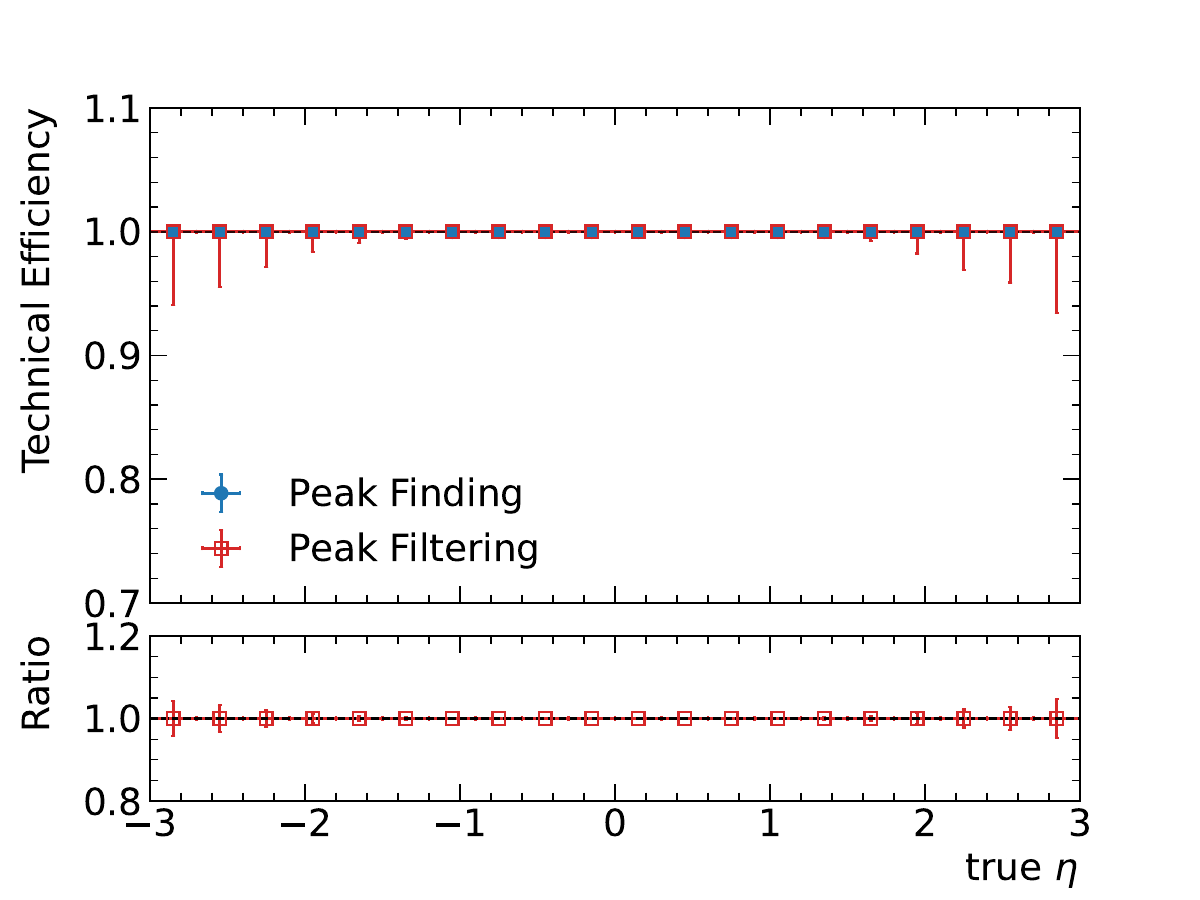}
    \end{subfigure}
    \hfill
    \begin{subfigure}{0.48\textwidth}
        \centering
        \includegraphics[width=\textwidth]{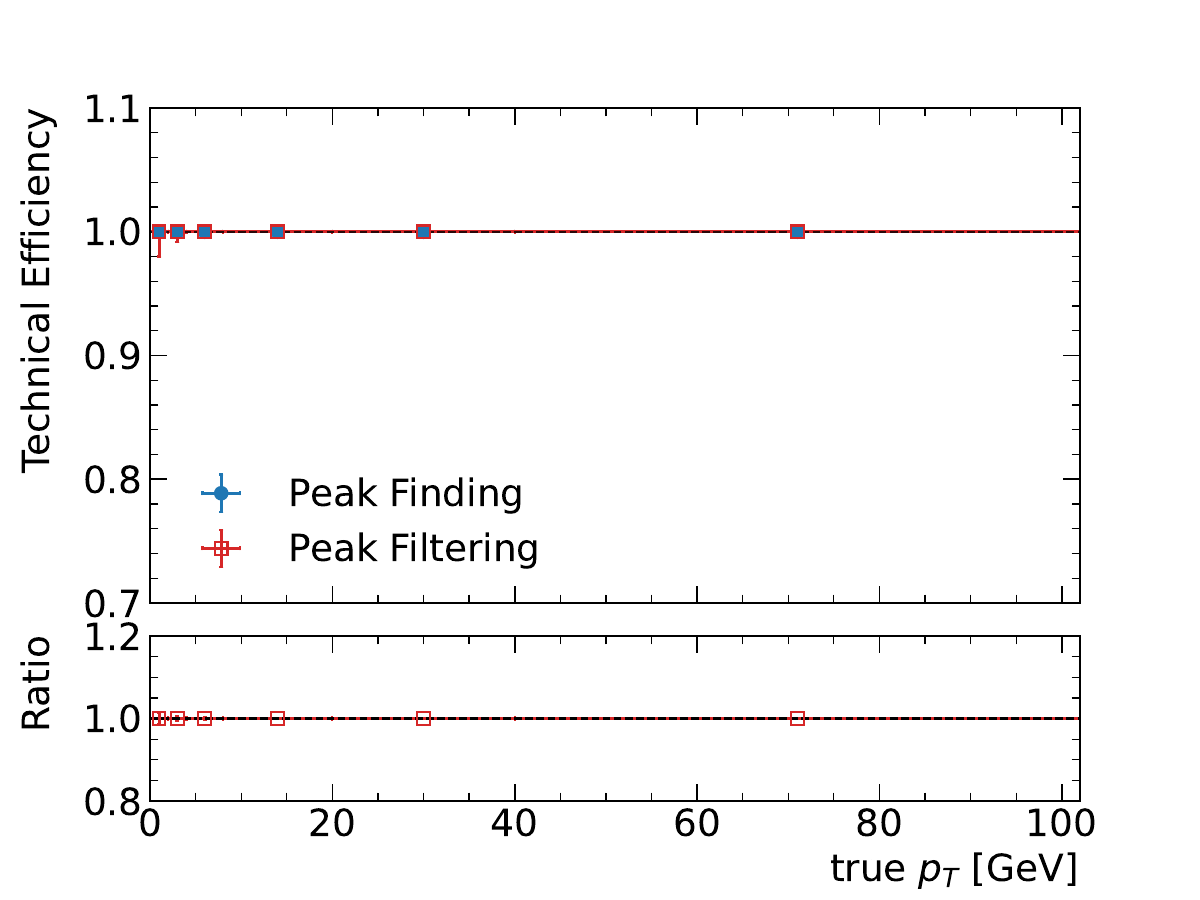}
    \end{subfigure}

     \begin{subfigure}{0.48\textwidth}
        \centering
        \includegraphics[width=\textwidth]{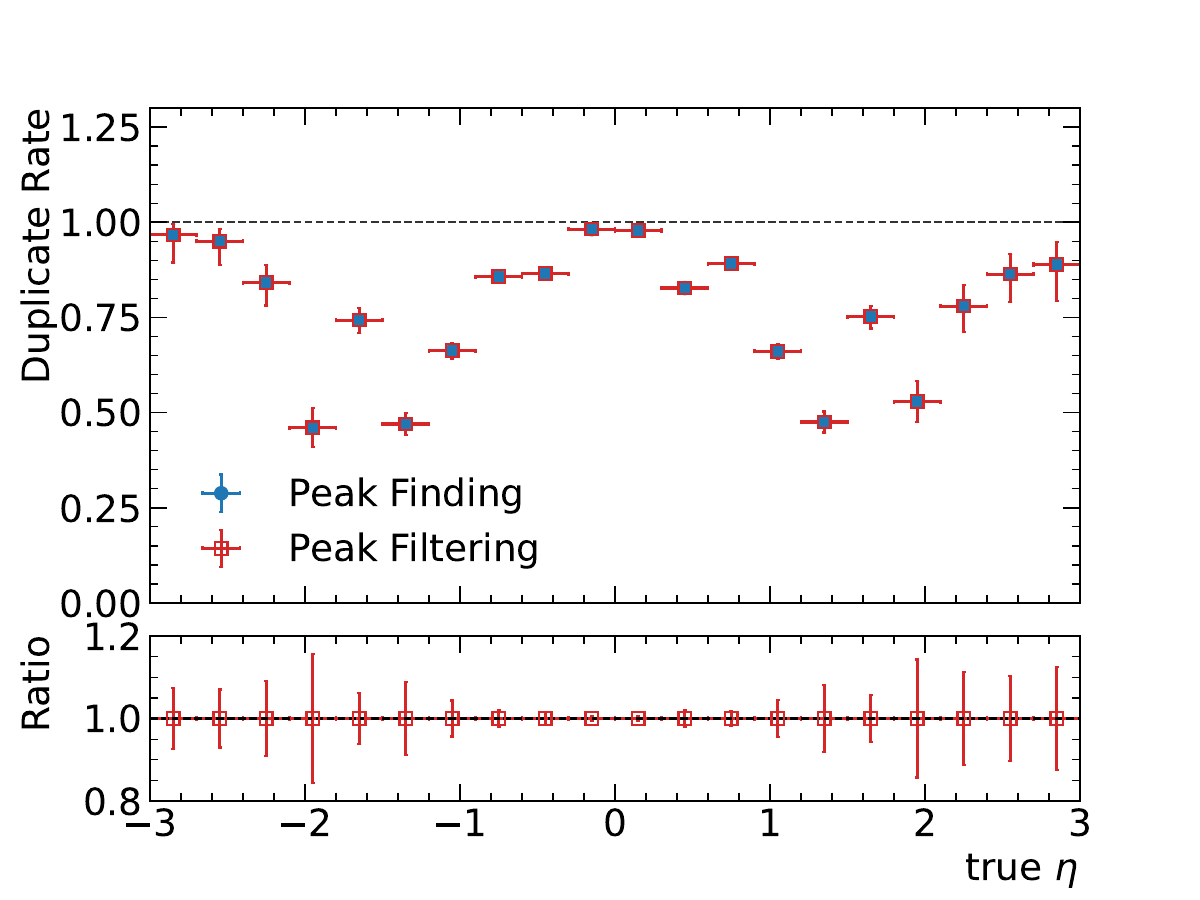}
    \end{subfigure}
    \hfill
    \begin{subfigure}{0.48\textwidth}
        \centering
        \includegraphics[width=\textwidth]{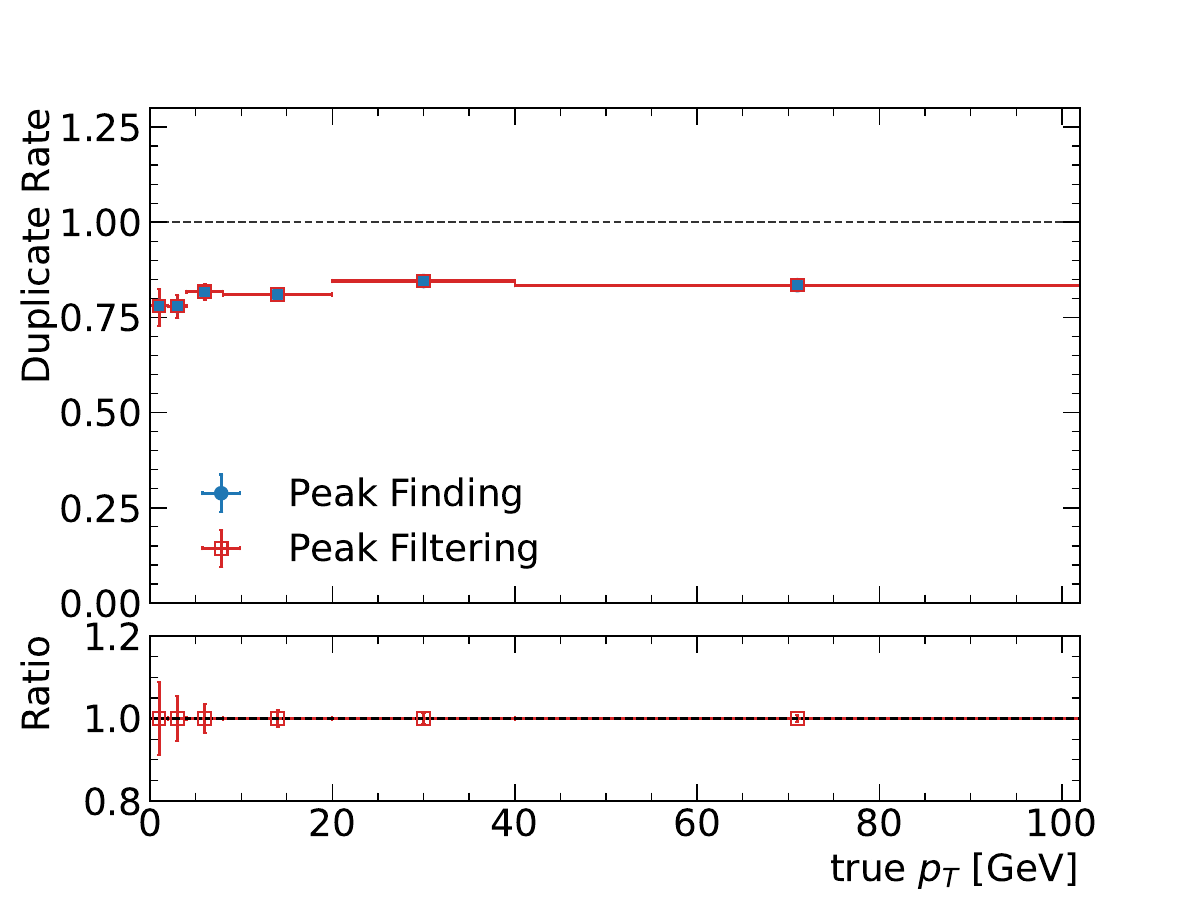}
    \end{subfigure}

    
    \caption{Seeding performance for a single muon $\langle \mu \rangle = 0$ sample. Technical efficiency and duplicate rate are shown as functions of the pseudo-rapidity and $p_T$ of the truth particle. 
    The blue circles show the performance of the maxima finding algorithm, while the red squares show the combined performance including neural network filtering. The bottom panels show the ratio between the filtering and the finding performance.}
    \label{fig:singleMuonPerformance}
\end{figure*}

For the characterization of the algorithm's performance, single muon events without \pileup were studied.
The results are shown in figure~\ref{fig:singleMuonPerformance}.
Overall, the results obtained in single-muon events without \pileup demonstrate excellent intrinsic performance of the algorithm.

For both the peak finding and filtering algorithms, the technical efficiency is found to be fully consistent with unity across the entire kinematic range considered in this study. No significant degradation is observed at low $p_T$ or in the forward detector regions, achieving optimal pattern-recognition performance under ideal detector conditions.
The algorithm shows a good degree of duplicate rate throughout the full phase space. The overlap between $\eta$ slices is reflected in the duplicate rate distribution.
The fake contribution is not shown in figure~\ref{fig:singleMuonPerformance} as no fake peak has been found over the entire considered parameter space.

\begin{figure*}
    \centering

    \begin{subfigure}{0.48\textwidth}
        \centering
        \includegraphics[width=\textwidth]{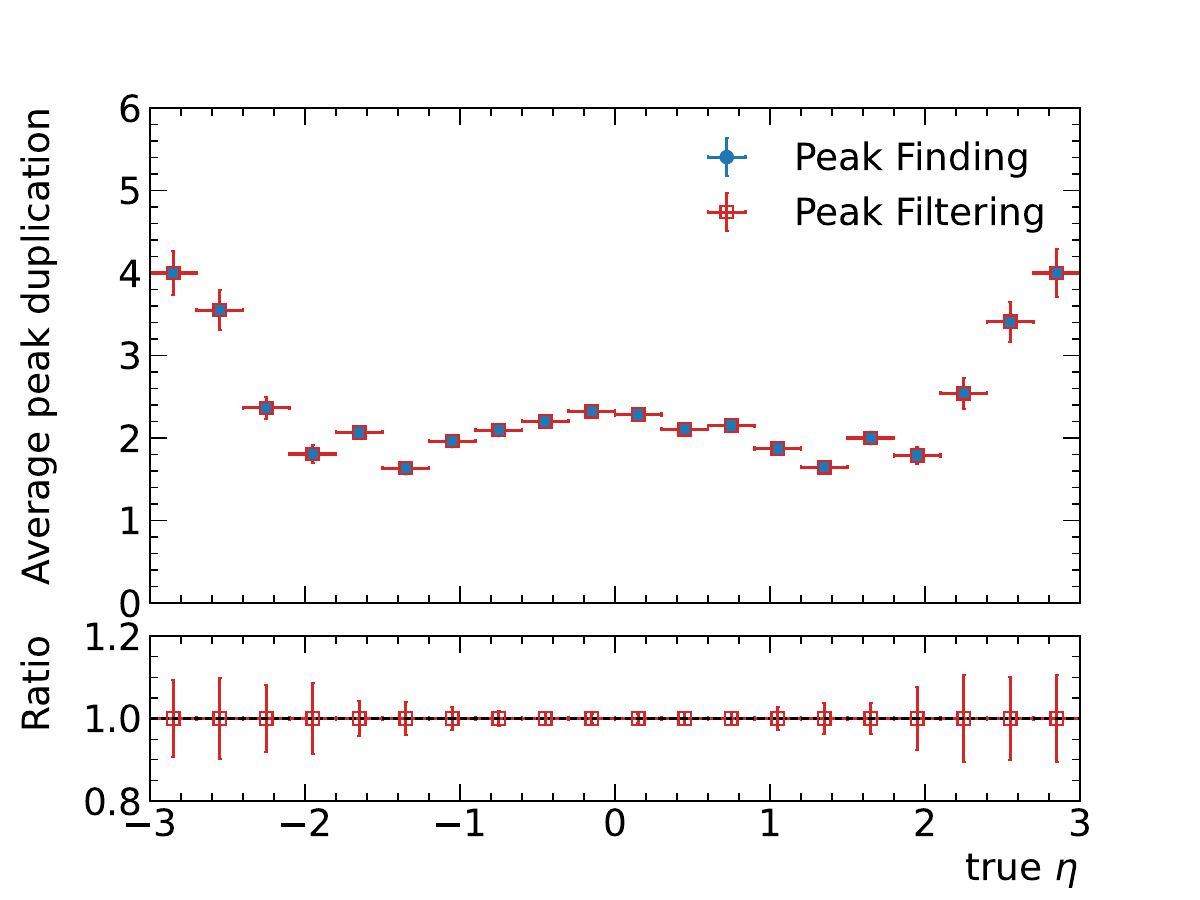}
    \end{subfigure}
    \hfill
    \begin{subfigure}{0.48\textwidth}
        \centering
        \includegraphics[width=\textwidth]{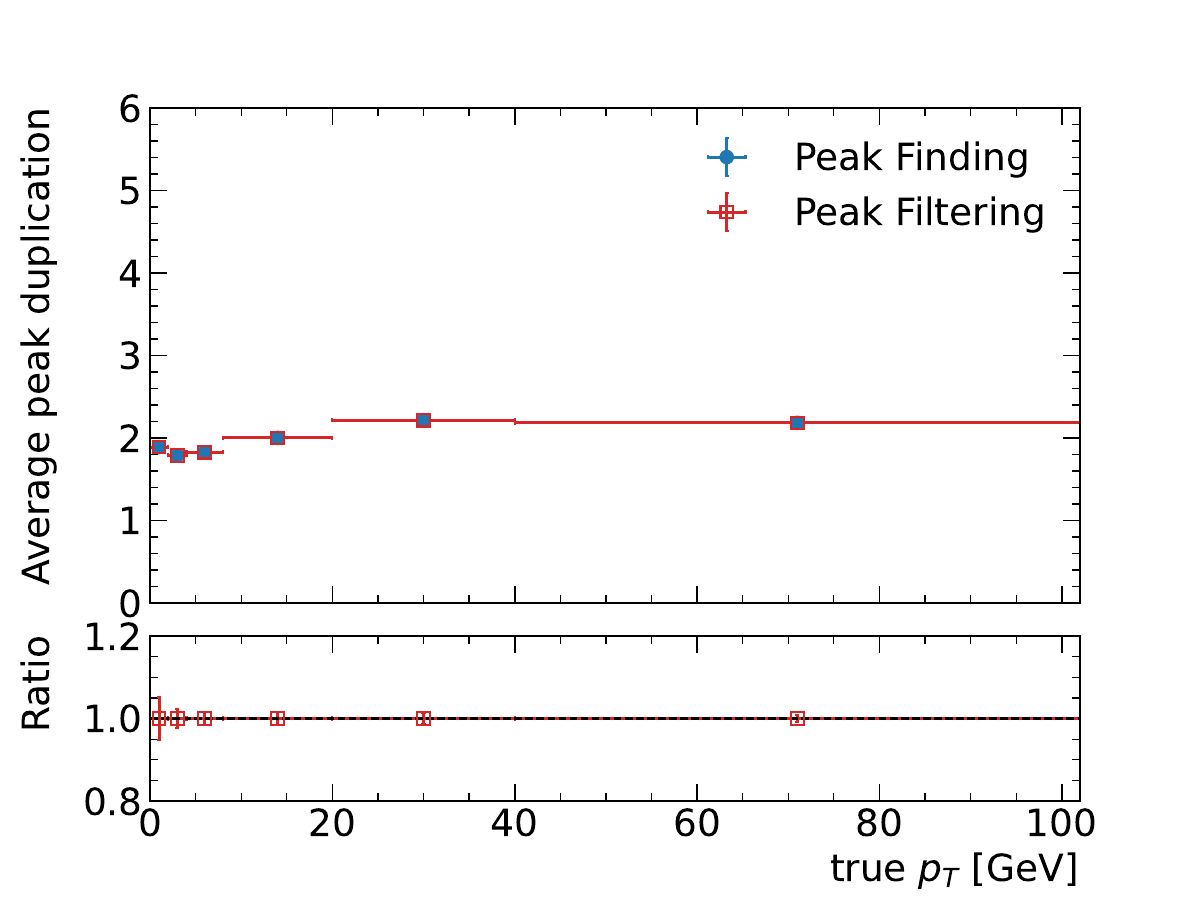}
    \end{subfigure}

    \caption{Average peak duplication per truth particle for the single muon $\langle\mu\rangle=0$ sample. The blue circles show performance of maxima finding step while red squares show combined performance including neural network filtering. The bottom panels show the ratio between the filtering and the finding performance.}
    \label{fig:singleMuonAverageDuplication}
\end{figure*}

Figure~\ref{fig:singleMuonAverageDuplication} presents the average number of peaks per truth particle as a function of $\eta$ and $p_T$. On average, two peaks are found for every truth particle, increasing to four at high $\eta$.

\subsection{Seeding performance in high \pileup conditions}
\label{sec:performance_in_high_pileup_conditions}

\begin{figure*}
    \centering

    \begin{subfigure}{0.48\textwidth}
        \centering
        \includegraphics[width=\textwidth]{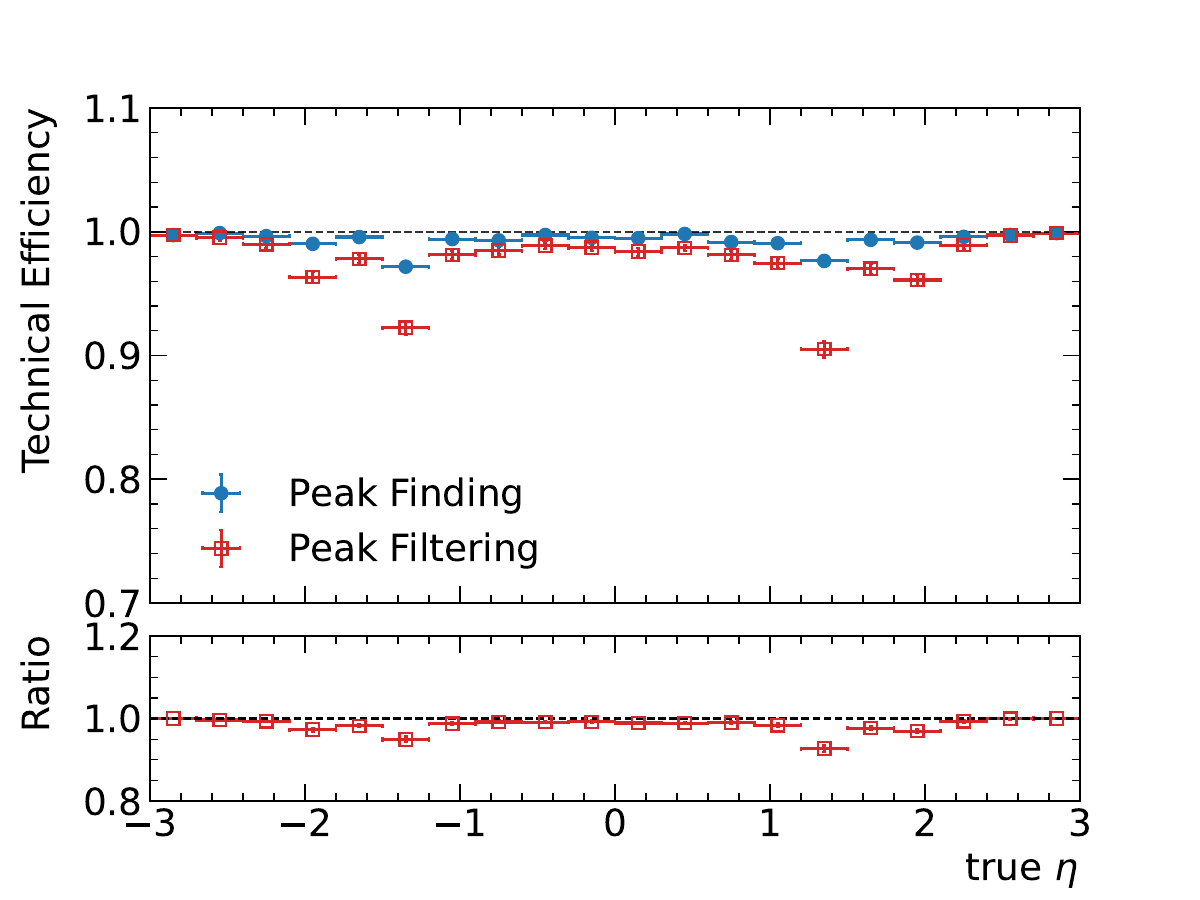}
    \end{subfigure}
    \hfill
    \begin{subfigure}{0.48\textwidth}
        \centering
        \includegraphics[width=\textwidth]{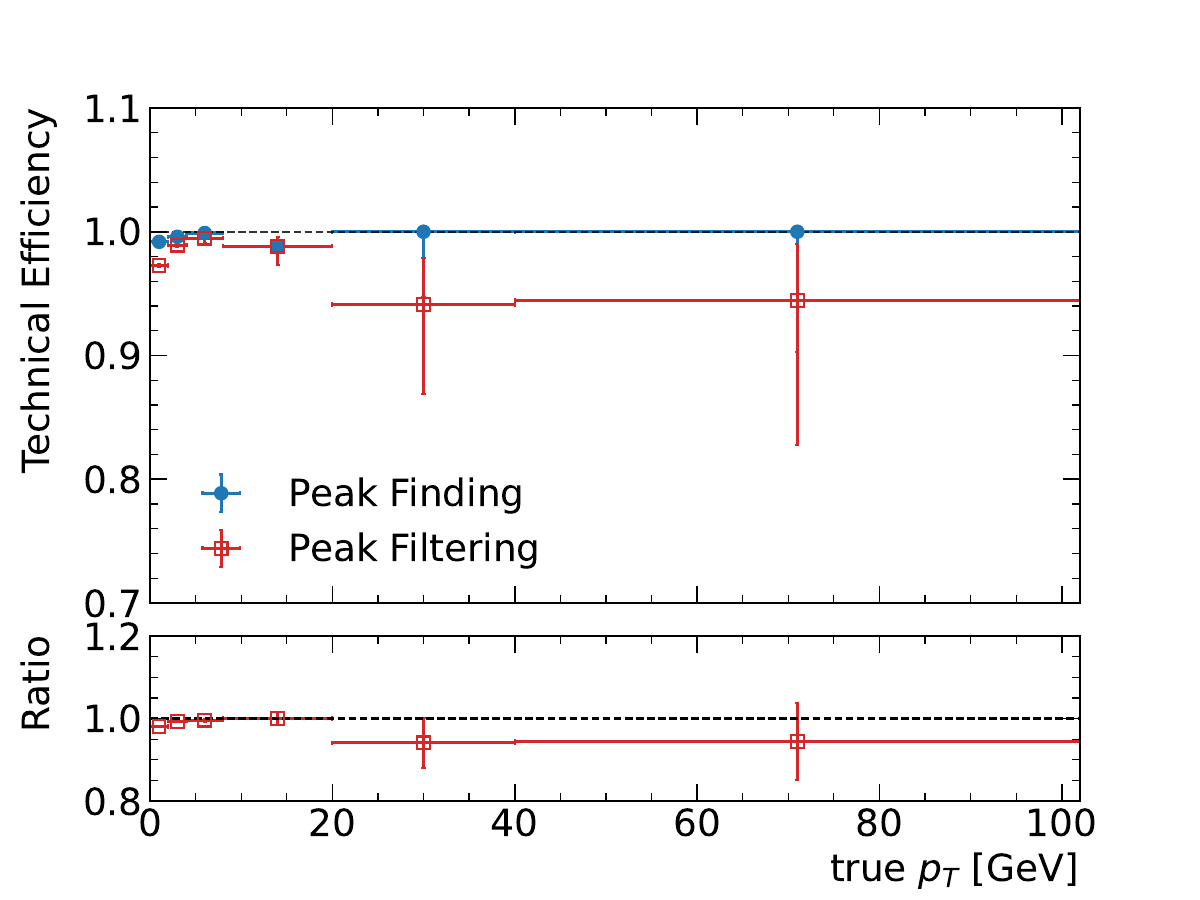}
    \end{subfigure}

     \begin{subfigure}{0.48\textwidth}
        \centering
        \includegraphics[width=\textwidth]{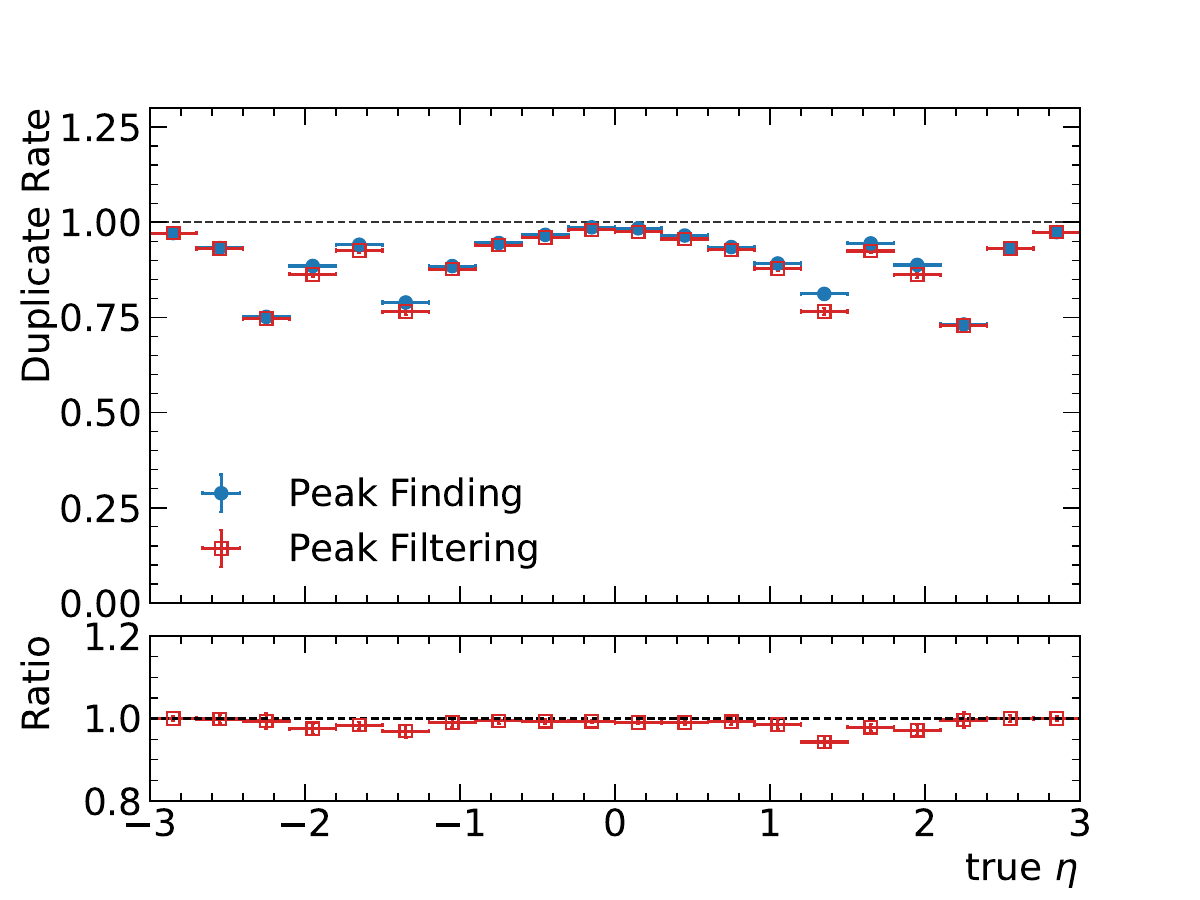}
    \end{subfigure}
    \hfill
    \begin{subfigure}{0.48\textwidth}
        \centering
        \includegraphics[width=\textwidth]{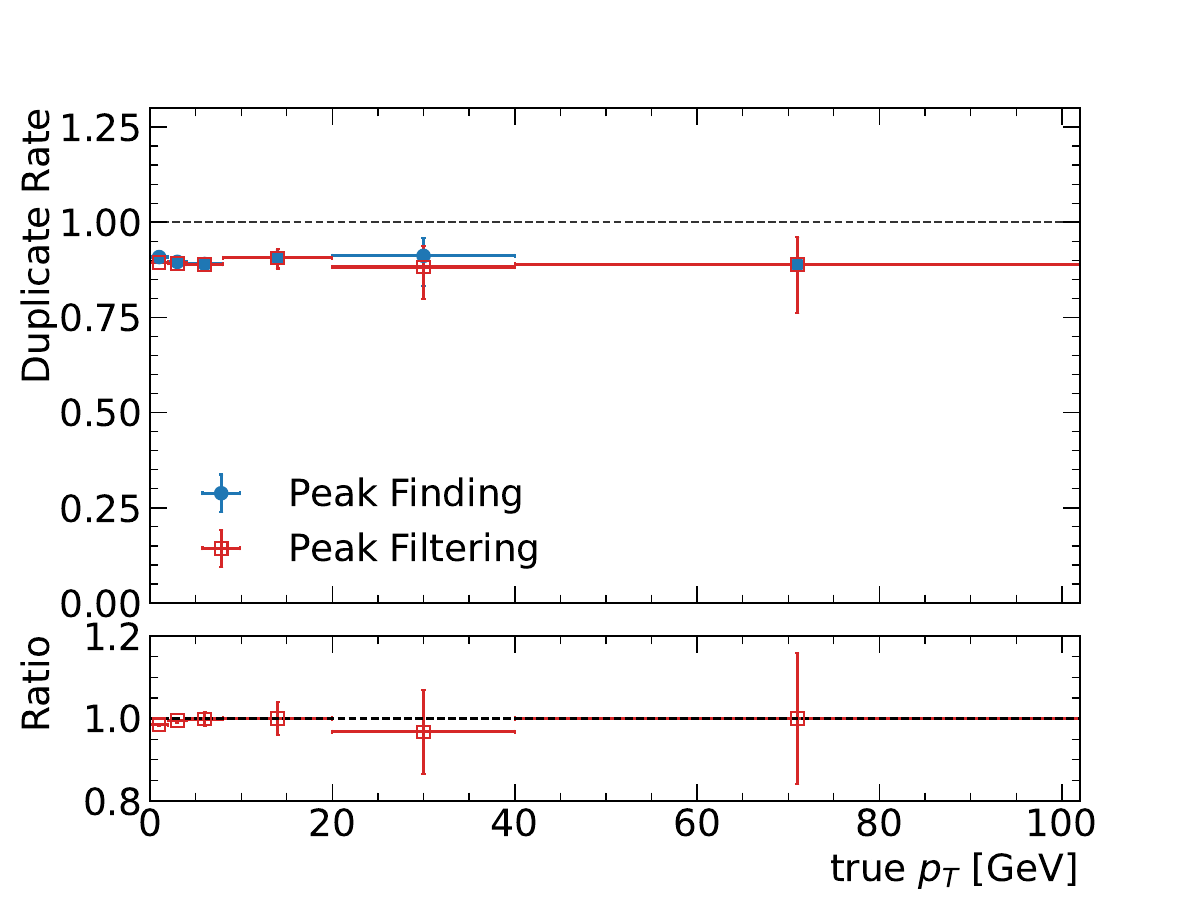}
    \end{subfigure}

    \begin{subfigure}{0.48\textwidth}
        \centering
        \includegraphics[width=\textwidth]{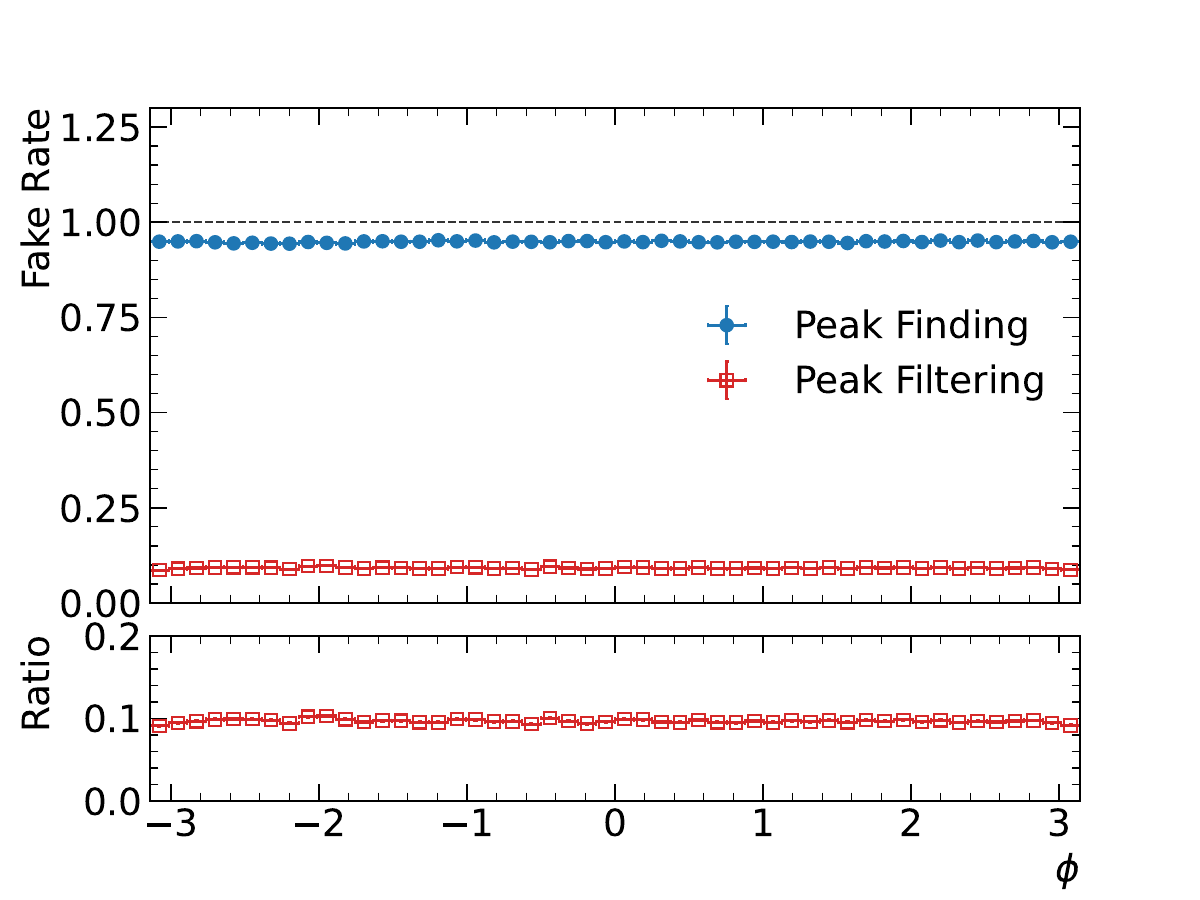}
    \end{subfigure}
    \hfill
    \begin{subfigure}{0.48\textwidth}
        \centering
        \includegraphics[width=\textwidth]{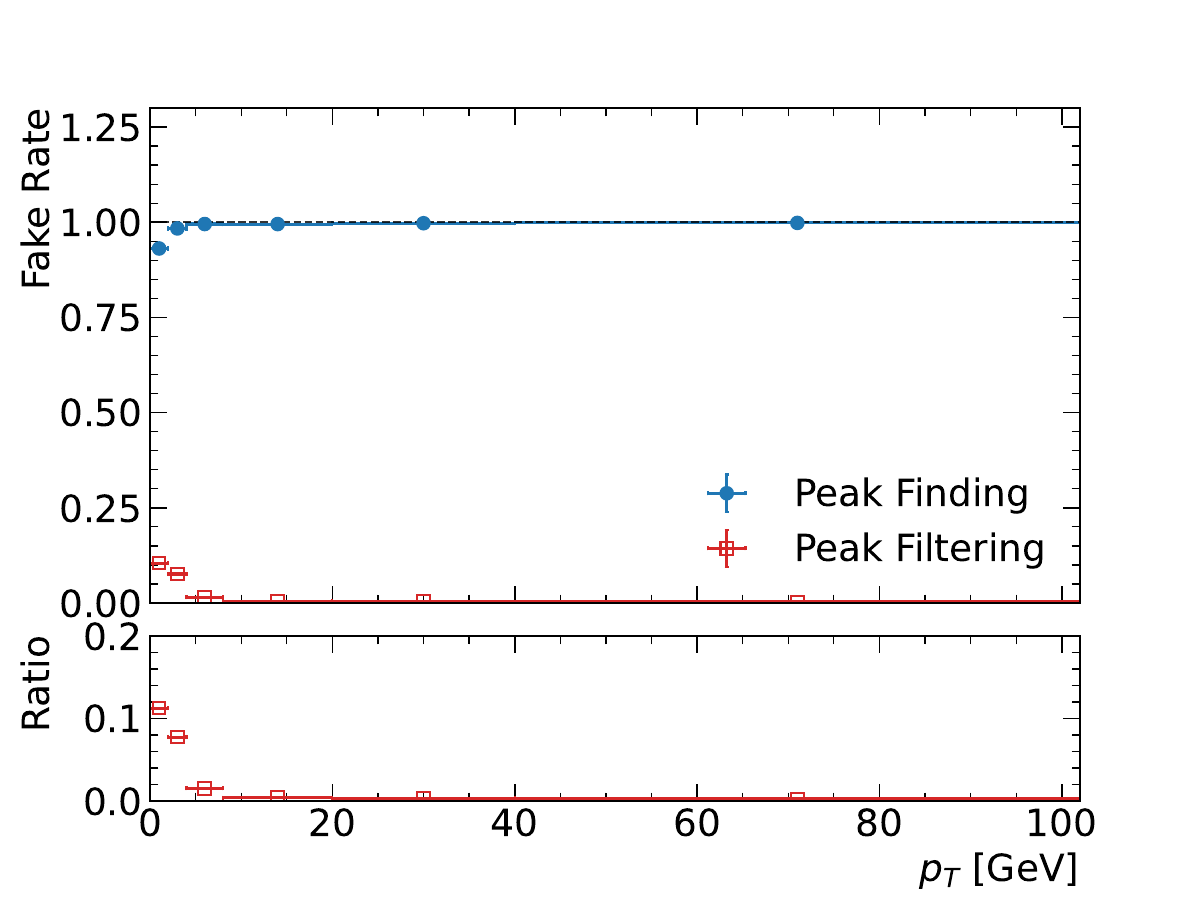}
    \end{subfigure}
    
    \caption{Seeding performance for a $t\bar{t}$ $\langle \mu \rangle = 200$ sample. Technical efficiency and duplicate rate are shown as functions of the pseudo-rapidity and $p_T$ of the truth particle. Fake rate is shown as a function of $\phi$ and $p_T$. The blue circles show the performance of the maxima finding algorithm, while the red squares show the combined performance including neural network filtering. The bottom panels show the ratio between the filtering and the finding performance.}
    \label{fig:ttBarPerformance}
\end{figure*}

For a realistic performance estimation of the discussed seeding algorithm, the $t\bar{t}$ events embedded in 200 inelastic \pileup pp events were studied. 
The results are shown in figure~\ref{fig:ttBarPerformance}.

The event environment is characterized by a very high detector occupancy, resulting in a large number of reconstructed peaks. In addition to the genuine peaks associated with truth particles, several additional peaks are observed in their immediate vicinity. These additional structures originate either from duplicate reconstructions of the same particle or from random combinations of unrelated hits produced by the high hit density. Table~\ref{tab:peak_numbers_ttBar} shows the average number per event of the peaks identified in each $\eta$ slice for both the peak finding and peak filtering stages. The number true peaks are also reported.

The peak-finding algorithm achieves an excellent efficiency across the full $\eta$ and $p_T$ range, with only a marginal loss of efficiency ($< 2 \%$) observed mainly for very low-$p_T$ tracks and in the $|\eta| \in [1, 2]$ region. The duplicate rate as a function of $\eta$ clearly reflects the overlap between adjacent $\eta$ slices, with enhanced duplicate production in the regions where neighboring $\eta$ sectors overlap. This demonstrates the robustness of the algorithm in maintaining high efficiency across sector boundaries.
The fake rate at the peak-finding stage is very large, reaching average values of approximately 95\%. This rate could be further reduced through additional optimization of the peak-finding algorithm.

The neural-network-based filtering stage significantly improves the purity of the reconstructed candidates while preserving the efficiency achieved by the peak-finding algorithm. Reduction in efficiency is observed in localized regions of $\eta$, especially in the interval corresponding to the transition between the strip barrel and end-cap detectors. A mild $p_T$ dependence of the efficiency is observed, with a marginal decrease in filtering efficiency at very low and at high $p_T$. 
The duplicate rate is only marginally affected by the neural-network filtering. In the context of the full tracking pipeline, these duplicates do not introduce a significant additional computational burden, as they are efficiently removed during the subsequent CKF stage.
A substantial improvement is observed in the fake rate, which is reduced to approximately 10\%, with the remaining fake contribution concentrated primarily in the low-$p_T$ region. For $p_T > 10$~GeV, the fake rate is compatible with zero within statistical uncertainties.

\begin{figure*}
    \centering

    \begin{subfigure}{0.48\textwidth}
        \centering
        \includegraphics[width=\textwidth]{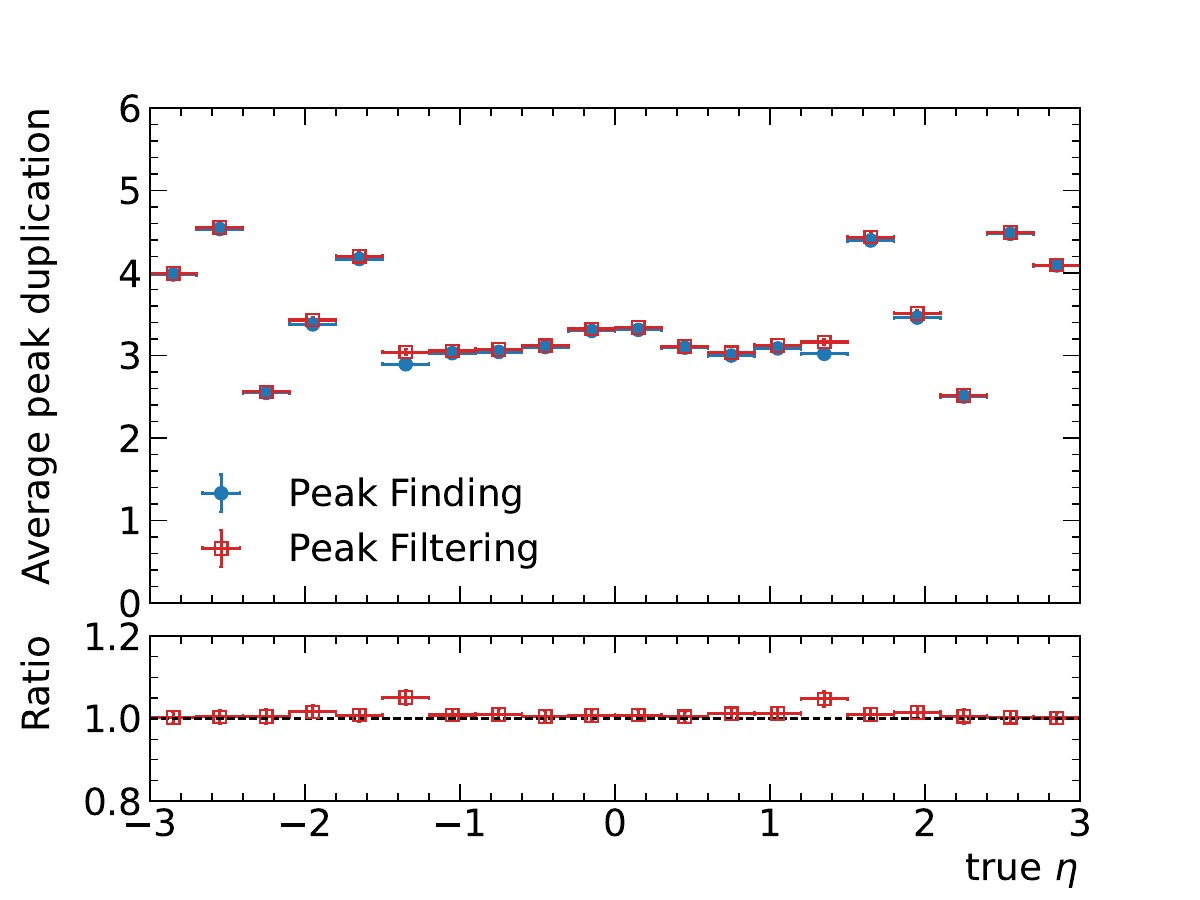}
    \end{subfigure}
    \hfill
    \begin{subfigure}{0.48\textwidth}
        \centering
        \includegraphics[width=\textwidth]{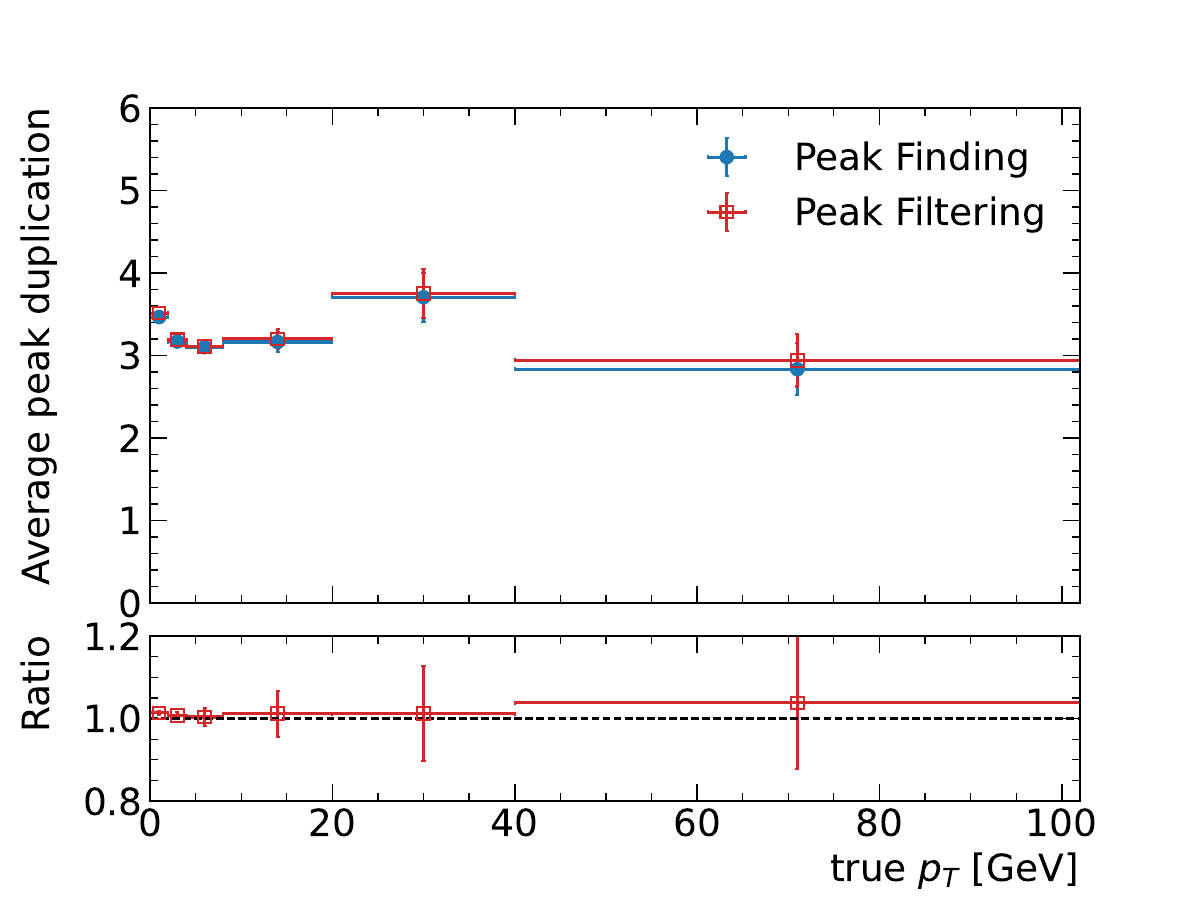}
    \end{subfigure}

    \caption{Average peak duplication per truth particle. The blue circles show performance of maxima finding step while red squares show combined performance including neural network filtering. The bottom panels show the ratio between the filtering and the finding performance.}
    \label{fig:ttBarAverageDuplication}
\end{figure*}

Figure~\ref{fig:ttBarAverageDuplication} presents the average number of peaks per truth particle as a function of $\eta$ and $p_T$. On average, three peaks are found for every truth particle.

\begin{table*}
\centering
\begin{tabular}{|c|c|c|c|c|}
\hline
\multirow{2}{*}[-0.7em]{\textbf{\# slice}}
& \multicolumn{2}{c|}{\textbf{Peak finding stage}}
& \multicolumn{2}{c|}{\textbf{Peak filtering stage}} \\
\cline{2-5}

 & 
\begin{tabular}[c]{@{}c@{}}\textbf{average number}\\\textbf{of total peaks}\end{tabular} &
\begin{tabular}[c]{@{}c@{}}\textbf{average number}\\\textbf{of true peaks}\end{tabular}  &
\begin{tabular}[c]{@{}c@{}}\textbf{average number}\\\textbf{of total peaks}\end{tabular} & 
\begin{tabular}[c]{@{}c@{}}\textbf{average number}\\\textbf{of true peaks}\end{tabular} \\
\hline
\textbf{0} & $1408.8 \pm 5.3$ & $311.1 \pm 2.5$ & $565.8 \pm 3.4$ & $298.7 \pm 2.4$ \\
\hline
\textbf{1} & $3583.7 \pm 8.5$ & $276.5 \pm 2.4$ & $641.0 \pm 3.6$ & $258.7 \pm 2.3$ \\
\hline
\textbf{2} & $17506.7 \pm 18.7$ & $342.6 \pm 2.6$ & $1302.7 \pm 5.1$ & $273.3 \pm 2.3$ \\
\hline
\textbf{3} & $8656.2 \pm 13.2$ & $301.4 \pm 2.5$ & $1088.8 \pm 4.7$ & $250.3 \pm 2.2$ \\
\hline
\textbf{4} & $2721.7 \pm 7.4$ & $251.1 \pm 2.2$ & $657.4 \pm 3.6$ & $229.7 \pm 2.1$ \\
\hline
\textbf{5} & $1826.9 \pm 6.0$ & $265.8 \pm 2.3$ & $556.8 \pm 3.3$ & $248.8 \pm 2.2$ \\
\hline
\textbf{6} & $1669.9 \pm 5.8$ & $270.4 \pm 2.3$ & $540.3 \pm 3.3$ & $254.6 \pm 2.3$ \\
\hline
\textbf{7} & $1725.1 \pm 5.9$ & $257.6 \pm 2.3$ & $535.0 \pm 3.3$ & $240.9 \pm 2.2$ \\
\hline
\textbf{8} & $2683.2 \pm 7.3$ & $248.2 \pm 2.2$ & $635.4 \pm 3.6$ & $223.9 \pm 2.1$ \\
\hline
\textbf{9} & $8899.5 \pm 13.3$ & $315.8 \pm 2.5$ & $1088.1 \pm 4.7$ & $247.3 \pm 2.2$ \\
\hline
\textbf{10} & $18133.5 \pm 19.0$ & $355.4 \pm 2.7$ & $1324.5 \pm 5.1$ & $276.2 \pm 2.4$ \\
\hline
\textbf{11} & $3432.7 \pm 8.3$ & $273.9 \pm 2.3$ & $628.5 \pm 3.5$ & $257.9 \pm 2.3$ \\
\hline
\textbf{12} & $1428.2 \pm 5.3$ & $316.5 \pm 2.5$ & $565.7 \pm 3.4$ & $304.3 \pm 2.5$ \\                 
\hline
\hline
\textbf{Total} & $73676.1 \pm 38.4$ & $3786.2 \pm 8.7$ & $10129.9 \pm 14.2$ & $3364.6 \pm 8.2$ \\                 
\hline
\end{tabular}
\caption{Average number of peaks per event obtained during the peak finding and the peak filtering stages. For each $\eta$ slice the average number of peaks and the average number of true peaks are reported for both stages. The total values, obtained by combining all the $\eta$ slices, are also included. These results are obtained from 50 $t\bar{t}$ events with $\langle\mu\rangle = 200$.}
\label{tab:peak_numbers_ttBar}
\end{table*}

\subsection{Effect of the neural network configuration on the peak filtering performance}
\label{sec:effect_neural_network_size_peak_identification_performance}

To optimize the peak-filtering performance and systematically explore the trade-off between model complexity, inference cost, and accuracy, a Bayesian hyper-parameter optimization algorithm~\cite{snoek2012practical} implemented in the Keras Tuner~\cite{omalley2019kerastuner} was employed. Instead of relying on arbitrary manual scaling of the baseline architecture, a broad scope of hyper-parameters and architectural topologies was evaluated. The optimization targeted the maximization of the Area Under the ROC Curve.

The search space included the number of convolutional blocks, the number of convolutions per block, base filter multiplicities, starting kernel sizes, activation functions, noise injection standard deviations, and regularization parameters such as dropout rates and weight decay. The optimization identified several highly efficient architectures, utilizing a Swish activation function rather than the baseline ReLU, and employing descending filter structures to minimize the parameter count without sacrificing representation power.

\begin{figure*}
    \centering

    \begin{subfigure}{0.48\textwidth}
        \centering
        \includegraphics[width=\textwidth]{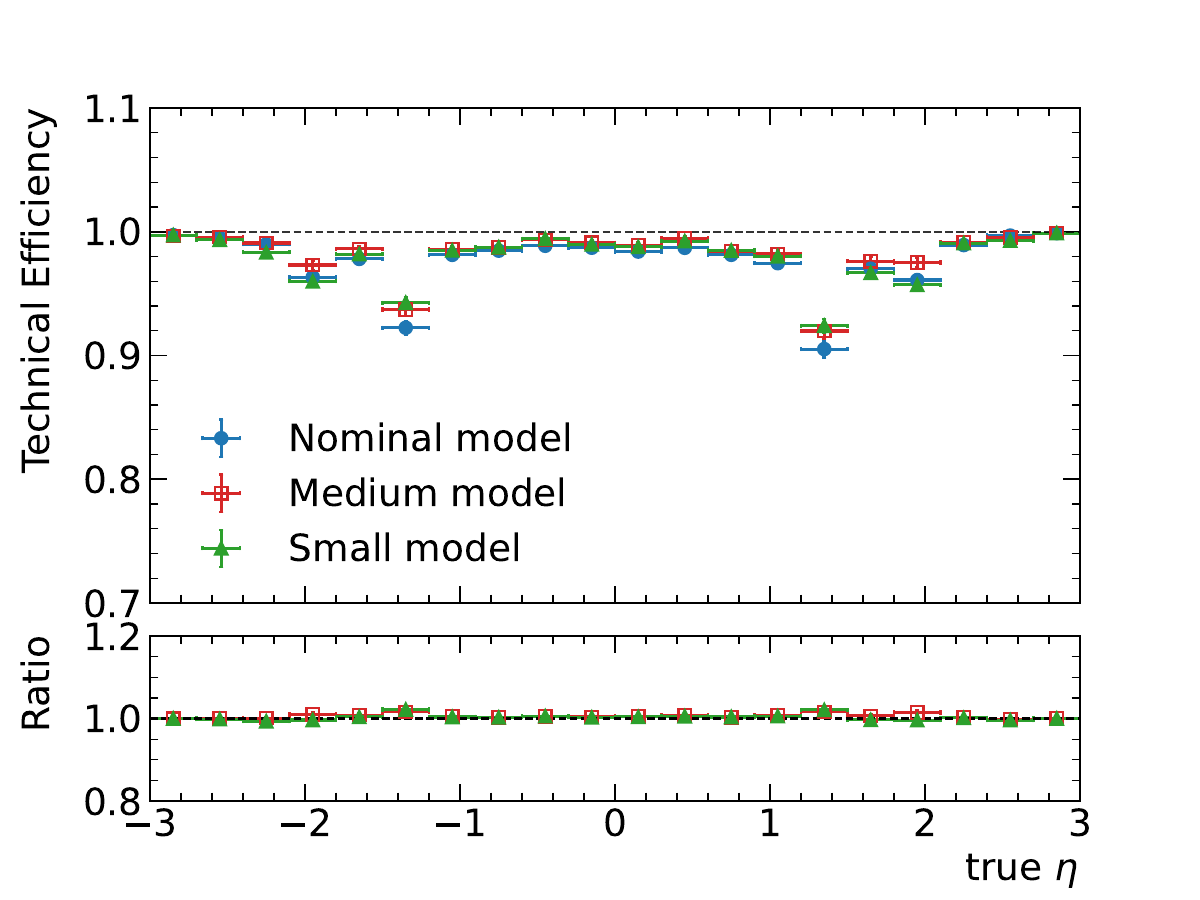}
    \end{subfigure}
    \hfill
    \begin{subfigure}{0.48\textwidth}
        \centering
        \includegraphics[width=\textwidth]{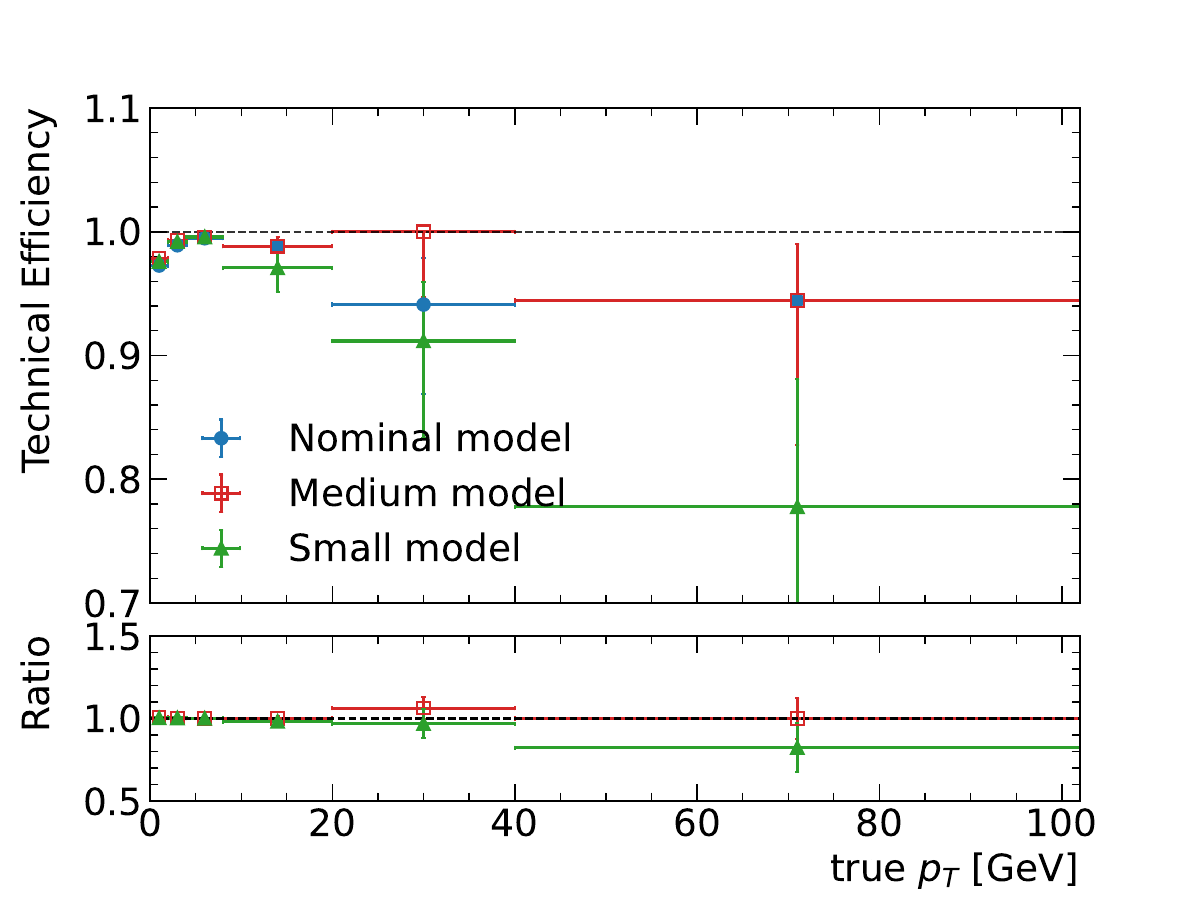}
    \end{subfigure}
    \begin{subfigure}{0.48\textwidth}
        \centering
        \includegraphics[width=\textwidth]{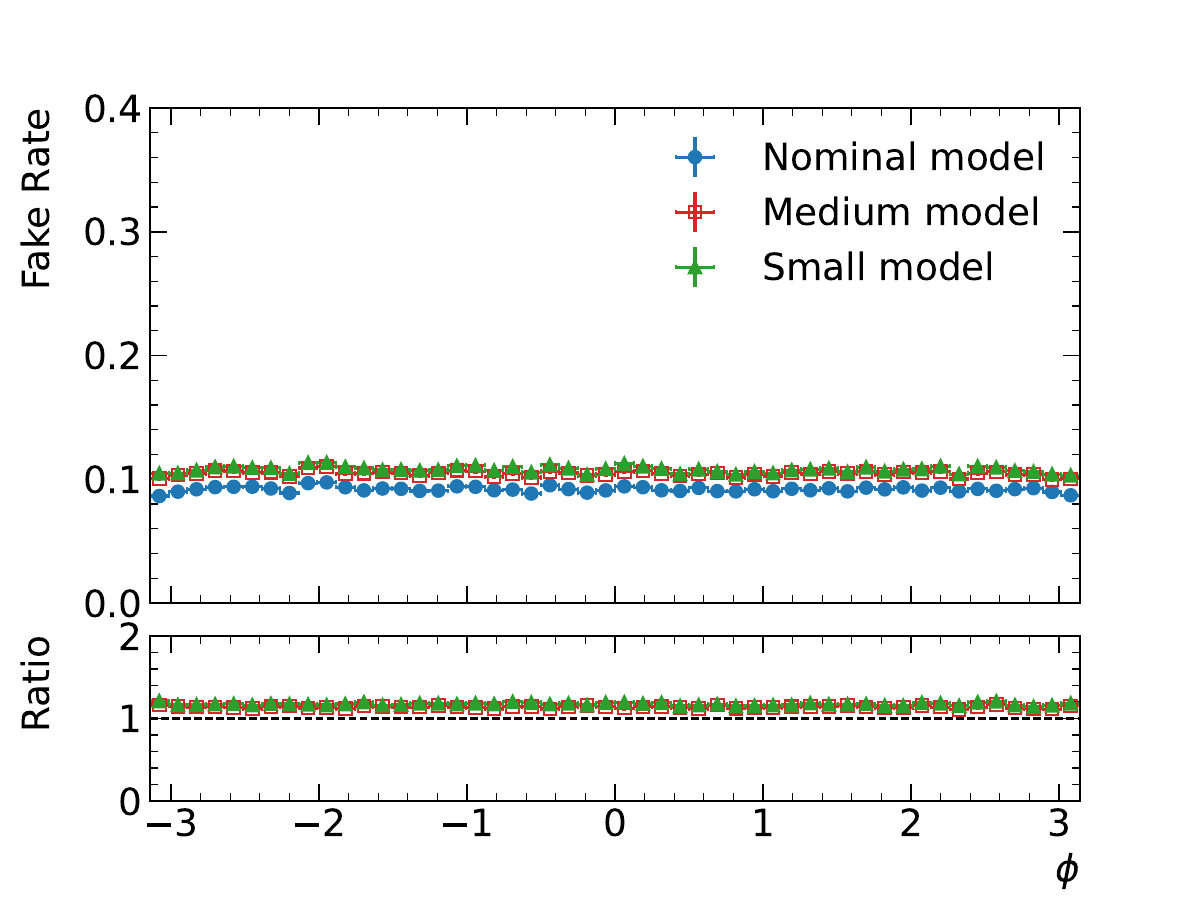}
    \end{subfigure}
    \hfill
    \begin{subfigure}{0.48\textwidth}
        \centering
        \includegraphics[width=\textwidth]{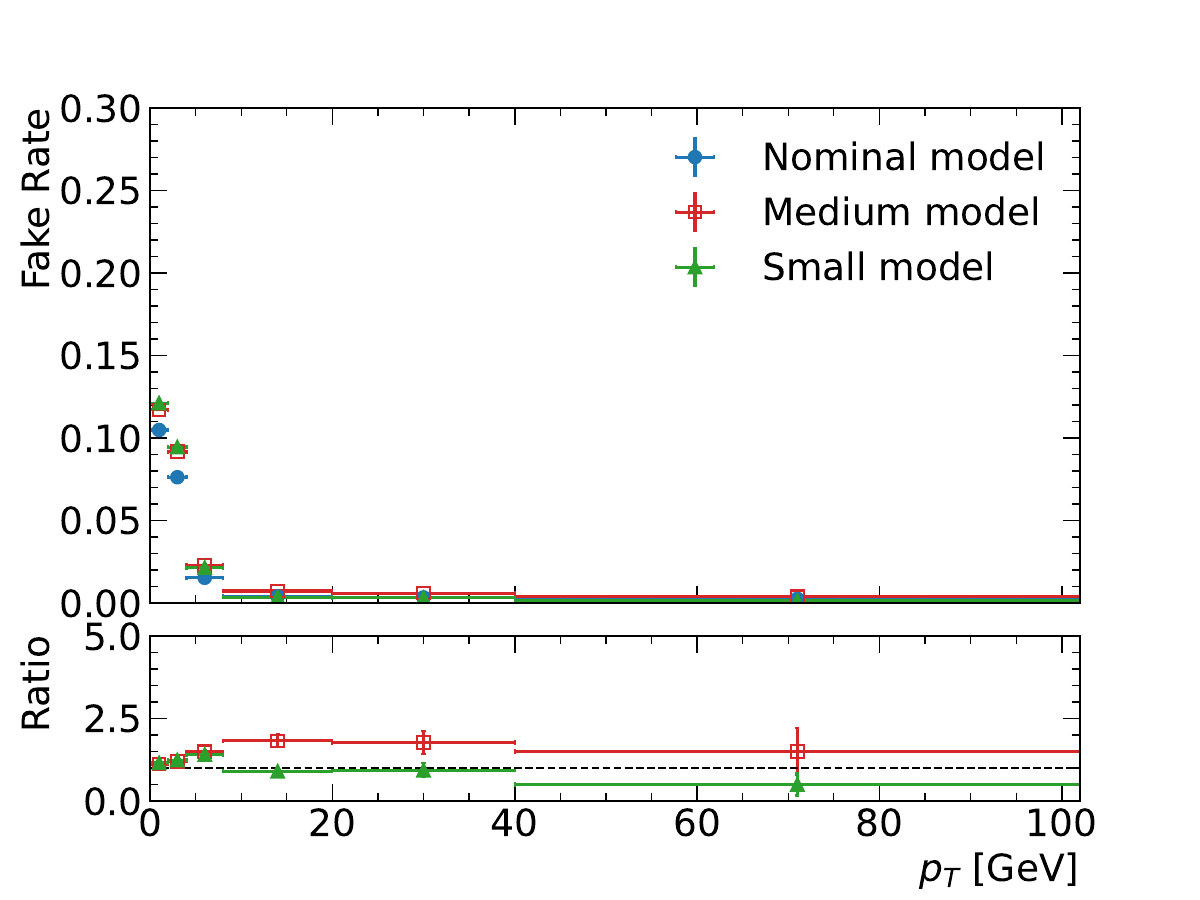}
    \end{subfigure}

    \caption{Peak filtering performance of the alternative neural networks. The nominal neural network model (blue) is compared to the alternative models, labeled as medium (red) and small (green) models. Efficiencies are shown on the top row, while fake rates are shown at the bottom. The bottom panels show the ratio with respect to the nominal performance.}
    \label{fig:neuralNetworkSizeImpact}
\end{figure*}

From this study, two optimized configurations were selected to be compared against the nominal model:
\begin{itemize}
    \item \textbf{Medium}: An intermediate model balancing performance and computational cost. It features five blocks starting with a $9 \times 9$ kernel and 32 filters. The architecture halves the number of filters at each successive block and applies targeted dropout in the early layers.
    \item \textbf{Small}: A highly compact network optimized for rapid inference. It maintains the five-block depth and initial $9 \times 9$ kernel of the medium model but restricts the base capacity to 16 filters. It uses the same filter-halving topology to minimize the parameter footprint while retaining robust discrimination.
\end{itemize}
The effect of these modifications on the peak-filtering efficiency and fake rate was evaluated using $t\bar{t}$ samples with $\langle\mu\rangle = 200$. The same composition of training and test samples were used to train the different models. The results are shown in figure~\ref{fig:neuralNetworkSizeImpact}. Since the selection on the neural-network discriminant score was kept fixed at a threshold of 0.5, each alternative configuration was trained and optimized independently within its respective architectural constraints.

The alternative trainings resulted in minimal or statistically insignificant changes in technical efficiency. The observed differences are concentrated around $|\eta| \simeq 2$ -- i.e. the detector transition region between the barrel and end-cap where the reconstructed peak distributions become noticeably noisier -- and at high $p_T$. However, a $\sim 20$\% increase in the fake rate is observed, which is the primary quantity the network is designed to suppress. In some regions of phase space the increase in the fake-candidate rate can be substantial.

These results indicate that network performance is sensitive not only to the overall model size but also to the architectural and training choices used in its construction. Reducing the model complexity leads to a degradation in discrimination power and an increased fake rate. A realistic optimization of the network design would therefore require a multi-objective approach, simultaneously considering efficiency, fake rate in different detector regions, inference cost, and model complexity. Such an optimization could be formulated through the extraction of a Pareto front~\cite{griffel2003multi}.

\subsection{Track reconstruction performance}
\label{sec:track_reconstruction_performance}

\begin{figure*}
    \centering

    \begin{subfigure}{0.48\textwidth}
        \centering
        \includegraphics[width=\textwidth]{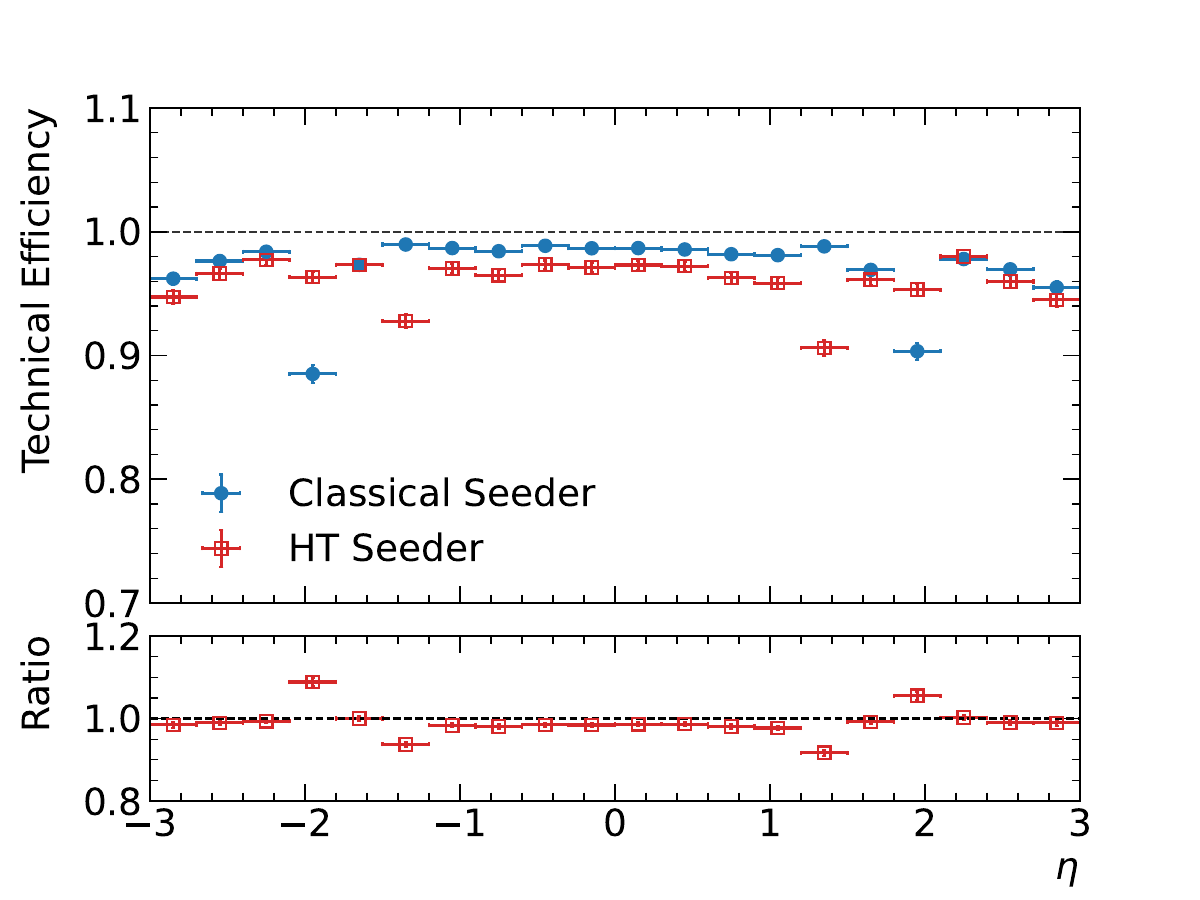}
    \end{subfigure}
    \hfill
    \begin{subfigure}{0.48\textwidth}
        \centering
        \includegraphics[width=\textwidth]{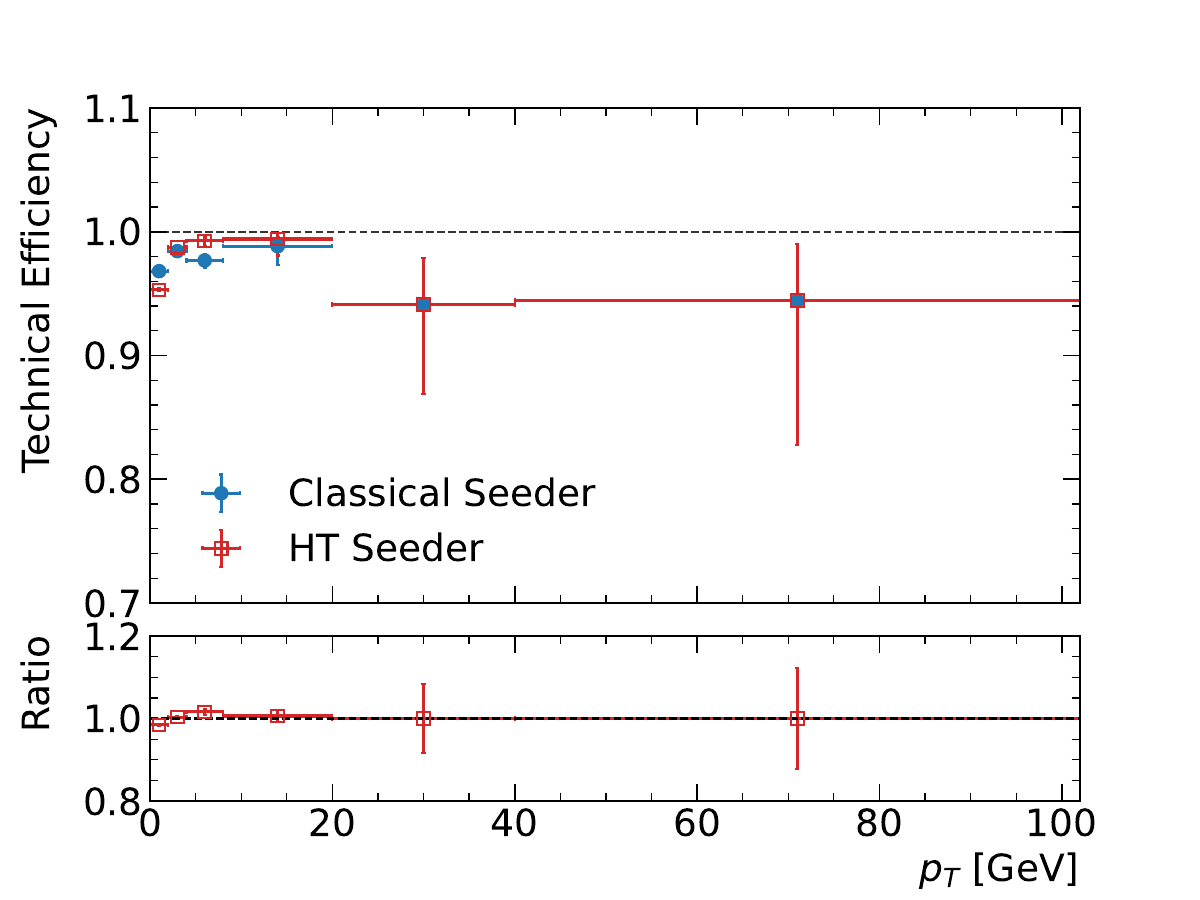}
    \end{subfigure}

    \caption{Track reconstruction technical efficiencies as a function of $\eta$ (left) and $p_T$ (right) of the truth particle for a $t\bar{t}$ \pileup 200 sample. 
    Tracks are obtained by running the Combinatorial Kalman Filter on the seed candidates. The seed candidates are produced by the ML-based seed filtering algorithm (red) or a classical triplet-finder seeder (blue). 
    The bottom panels show the ratio between the two workflow's performance.
    }
    \label{fig:ttBarTrackPerformance}
\end{figure*}

While the focus of this study is the stand-alone seeding performance, which provides an important characterization of the algorithm's behavior, an additional validation must be performed within the context of the full track-reconstruction chain. The relevant figure of merit is the impact of the new seeding algorithm on the technical efficiency of the downstream tracking stages. To assess this effect, a simulated $t\bar{t}$ sample with an average \pileup of 200 interactions was used.


The seeds produced by the HT-based algorithm were used as input to the CKF, after undergoing an additional cleaning step. Although the Hough Transform efficiently groups measurements belonging to the same charged particle, a small fraction of outlier space points can still be associated with a seed because of combinatorial ambiguities. Such outliers can significantly bias the estimation of the initial track parameters required by the CKF, degrading the track finding performance. To mitigate this effect, a dedicated cleaning procedure was developed. The algorithm exploits the consistency of the $\cot(\theta)$ values computed from pairs of space points within each seed to identify incompatible measurements. Space points whose $\cot(\theta)$ is inconsistent with the majority of the seed are classified as outliers and removed. The cleaned seeds are then used as input to the CKF track finding algorithm.

For comparison, the same CKF configuration was also run using seeds produced by the triplet-based seeding algorithm available in \ACTS -- here referred to as "classical seeder" --, which constructs seeds from triplets of space points compatible with helical trajectories. Apart from the seeding stage, the two reconstruction chains are identical, including the CKF configuration and all downstream reconstruction steps. Consequently, any differences observed in the reconstructed track performance can be directly attributed to the different seeding strategies. The resulting tracking performance is shown in Fig.~\ref{fig:ttBarTrackPerformance}.

It should be noted that the CKF configuration has not been specifically tuned for this seed topology; therefore, the achieved performance may not be optimal.
Despite this limitation, the technical tracking efficiency as a function of both $\eta$ and $p_T$ demonstrates good performance across the full kinematic range considered in this study. The track reconstruction technical efficiency is stable and uniform across the kinematic phase space considered, remaining above 90\% throughout the entire phase space considered. Degradation of performance can be noticed, especially at low $p_T$, with technical efficiency reaching 96\%.

%% file: sections/parallelization_capabilities.tex
The algorithm admits several opportunities for parallelization at multiple levels, making it well suited for execution on massively parallel architectures such as GPUs. 

The primary source of parallelism arises from the partitioning of the detector into $\eta$ slices, which can be processed independently since no information exchange is required during either the seeding or track-finding stages. Thus, multiple instances of both the HT-based seeding and the CKF can be executed concurrently.

In addition, several stages of the reconstruction chain can be parallelized within a single slice. Under the assumption that event space point positions are already resident in GPU memory, the reconstruction of track candidates can be formulated as a sequence of computational kernels.
The association between space points and detector slices can be implemented by a kernel operating on individual space points, with the kernel evaluating for every space point its compatibility with the geometric definition of all slices. This would result in the production of a slice bit mask for each space point, encoding its membership across slices.
To improve memory locality -- thus more efficient downstream processing --, the space points can subsequently be reordered into a contiguous memory layout. 
Following this step, the HT space will be populated by assigning kernels to process disjoint fragments of the parameter space.
A natural mapping of computational tasks to the HT space can be achieved by assigning one kernel to each column, corresponding to a fixed value of $q/p_T$. This mapping ensures a regular workload distribution and facilitates efficient parallel execution. 
The search for local maxima in the HT space can subsequently be parallelized by partitioning the parameter space into an appropriate number of kernels. This stage may be decomposed into maxima identification, maxima counting, and extraction of candidate maxima coordinates.
Finally, the regions surrounding the identified maxima can be processed independently by dedicated kernels. These kernels would perform inference procedures and subsequently classify the corresponding maxima, flagging those that satisfy the selection criteria as valid track candidates.

%% file: sections/summary.tex
The paper presents a hybrid method for particle track seeding that combines the \Hough transform  with neural network–based image recognition. 
The approach leverages the strengths of the \Hough Transform for initial candidate generation while addressing its well-known tendency to produce a large number of fake candidates through a subsequent neural network filtering stage.

A key innovation lies in reusing the HT representation of track candidates in the $(q/p_T, \phi)$ parameter space as input to the neural network. 
This design choice, though conceptually simple, distinguishes the method from prior work and enables a more coherent processing pipeline. 
The approach minimizes computationally expensive transformations between data formats by maintaining a consistent representation across both the peak-finding and neural filtering stages.

The algorithm is designed with parallelization in mind, making it well-suited for future implementation on computing accelerators such as GPUs, although such deployment has not yet been realized. 
Physics performance has been evaluated under realistic High-Luminosity LHC conditions using the \ACTS toolkit and the Open Data Detector. 
Results demonstrate high efficiency alongside a significantly reduced rate of fake candidates.

Future work will focus on evaluating the computational performance of this algorithm, implementing the algorithm on GPU architectures, integrating it more fully within the \ACTS framework, and exploring the use of neural networks earlier in the pipeline, particularly during peak finding in HT space. 
Additional studies will be required to assess robustness under realistic detector imperfections and to evaluate performance for challenging cases such as electron track reconstruction, ensuring the method’s suitability for deployment in experimental environments.

%% file: sections/ai.tex
During the preparation of this work, the authors used ChatGTP and Writefull for linguistic editing, including improvements to grammar, style, and clarity. After using this tool/service, the authors reviewed and edited the content as needed and take full responsibility for the content of the published article.

%% file: sections/acknowledgements.tex
The work of C. Varni was supported by the \emph{"Initiative of Excellence -- Research University"} program at AGH University of Krakow. The work of K. Cie\'sla, T. Bo\l d and M. Wolter was supported by the National Science Center (NCN) of Poland under research project UMO-2023/51/B/ST2/00920.